# Scheduling of Separable Mobile Energy Storage Systems with Mobile Generators and Fuel Tankers to Boost Distribution System Resilience

Wei Wang, *Student Member, IEEE*, Xiaofu Xiong, *Member, IEEE*, Yufei He, Jian Hu, Hongzhou Chen.

***Abstract*—Mobile energy resources (MERs) have been shown to boost DS resilience effectively in recent years. In this paper, we propose a novel idea, the separable mobile energy storage system (SMESS), as an attempt to further extend the flexibility of MER applications. "Separable" denotes that the carrier and the energy storage modules are treated as independent parts, which allows the carrier to carry multiple modules and scatter them independently throughout the DS. The constraints for scheduling SMESSs involving carriers and modules are derived based upon the interactive behavior among them and the DS. In addition, the fuel delivery issue of feeding mobile emergency generators (MEGs), which was usually bypassed in previous studies involving the scheduling of MEGs, is also considered and modeled. SMESSs, MEGs, and fuel tankers (FTs) are then jointly routed and scheduled, along with the dynamic DS reconfiguration, for DS service restoration by integrating them in a mixed-integer linear programming (MILP) model. Finally, the test is conducted on a modified IEEE 33-node test system, and results verify the effectiveness of the model in boosting DS resilience.**

## NOMENCLATURE

*Sets*

| | |
|---|---|
| $\mathcal{T}$ | Set of time spans. $\mathcal{T}$={1, 2, …, $D$}, where $D$ is the length of scheduling horizon. |
| $\mathcal{N}/\mathcal{N}_{\text{DP}}$ | Set of nodes/fuel depots. |
| $\mathcal{N}_{\text{M}}/\mathcal{N}_{\text{S}}/\mathcal{N}_{\text{G}}$ | Set of nodes supporting the access of MERs/SMESSs/MEGs. |
| $\mathcal{M}/\mathcal{M}_{\text{S}}/\mathcal{M}_{\text{G}}/\mathcal{M}_{\text{F}}$ | Set of MERs/Carrs (of SMESSs)/MEGs/FTs. |
| $\mathcal{K}$ | Set of Mods of SMESSs. |
| $\mathcal{L}$ | Set of DS branches. |
| $\mathcal{N}_{\text{FO},t}/\mathcal{N}_{\text{FC},t}$ | Set of nodes with faulted open/closed load switches during time span $t$. |
| $\mathcal{L}_{\text{FO},t}/\mathcal{L}_{\text{FC},t}$ | Set of faulted open/closed branches during time span $t$. |

*Variables*

| | |
|---|---|
| $x_{j,i,t}$ | Binary variable; 1 if MER $j$ is parked at node $i$ during time span $t$, 0 otherwise. |
| $v_{j,i,t}$ | Binary variable; 1 if MER $j$ is traveling to node $i$ during time span $t$, 0 otherwise. |
| $S_{j,t}$ | Travel time to be consumed by MER $j$ during time span $t$. |
| $R_{j,t}$ | Residual travel time of MER $j$ during time span $t$. |
| $\omega_{j,t}$ | Binary variable; 1 if MER $j$ is traveling during time spans $t$−1 and $t$. |
| $\zeta_{k,i,t}$ | Binary variable; 1 if Mod $k$ belongs to node $i$ during time span $t$, 0 otherwise. |
| $\gamma_{k,j,t}$ | Binary variable; 1 if Mod $k$ belongs to Carr $j$ during time span $t$, 0 otherwise. |
| $\alpha_{j,i,k,t}$ | Binary variable; 1 if Carr $j$ carrying Mod $k$ arrives at node $i$ during time span $t$, 0 otherwise. |
| $c_{k,i,t}/d_{k,i,t}$ | Binary variable; 1 if Mod $k$ charges/discharges at node $i$ during time span $t$, 0 otherwise. |
| $P^{\text{c.S}}_{k,i,t}/P^{\text{d.S}}_{k,i,t}$ | Active power output of Mod $k$ that charges/discharges at node $i$ during time span $t$. |
| $Q^{\text{S}}_{k,i,t}$ | Reactive power output of Mod $k$ that charges/discharges at node $i$ during time span $t$. |
| $SOC_{k,t}$ | State of charge of Mod $k$ at the end of time span $t$. |
| $P^{\text{G}}_{m,i,t}/Q^{\text{G}}_{m,i,t}$ | Active/reactive power output of MEG $m$ at node $i$ during time span $t$. |
| $B_{m,i,t}/B^{+}_{m,i,t}$ | Gross/extra fuel demand for the generation of MEG $m$ at node $i$ during time span $t$. |
| $\tau_{m,t,l}$ | Binary variable; 1 if the output of MER $m$ during time span $t$ is within the $l$th interval [$p_{m,l-1}$, $p_{m,l}$], 0 otherwise. |
| $SOF_{m/h/i,t}$ | State of fuel of MEG $m$/FT $h$/node (depot) $i$ at the end of time span $t$. |
| $b_{m,t}$ | Binary variable; 1 if $B_{m,i,t} \leq SOF_{m,t} \cdot F_m$, 0 otherwise. |
| $\iota_{m/h,i,t}$ | Binary variable; 1 if MEG m/FT h can exchange fuel with node (depot) $i$, 0 otherwise. |
| $G_{m,i,t}$ | Supplemented fuel of MEG $m$ at node (or fuel depot) $i$ during time span $t$. |
| $D_{h,i,t}$ | Fuel output of FT $h$ at node (or fuel depot) $i$ during time span $t$. |
| $f^{i_x}_{ii',t}$ | Flow of commodity $i_x$ from nodes $i$ to $i'$ during time span $t$. |
| $\lambda_{ii',t}$ | Binary variable; 1 if arc ($i$, $i'$) is included in the directed fictitious spanning tree during time span $t$, 0 otherwise. |
| $\mu_{ii',t}$ | Binary variable; 1 if branch ($i$, $i'$) is included in the fictitious spanning tree during time span $t$, 0 otherwise. |
| $\kappa_{ii',t}$ | Binary variable; 1 if branch ($i$, $i'$) of the DS is closed during time span $t$, 0 otherwise. |
| $P_{i'i,t}/Q_{i'i,t}$ | Active/reactive power flow on branch ($i'$, $i$) from node $i'$ to node $i$ during time span $t$. |
| $P^{\mathrm{IN}}_{i,t}/Q^{\mathrm{IN}}_{i,t}$ | Active/reactive power injected by a Mod or MEG into node $i$ during time span $t$. |
| $\delta_{i,t}$ | Binary variable; 1 if the load at node $i$ is picked up during time span $t$, 0 otherwise. |
| $V^2_{i,t}$ | Squared voltage magnitude at node $i$ during time span $t$. |
| $\eta_{i,t}$ | Binary variable; 1 if node $i$ is energized during time span $t$, 0 otherwise. |
| $\rho_{i,t}$ | Binary variable; 1 if at least one Mod or MEG is connected to node $i$ during time span $t$, 0 otherwise. |
| $\sigma_{i,t}$ | Binary variable; 1 if at least one energized node is connected to node $i$ during time span $t$, 0 otherwise. |
| $\chi_{i'\text{-}ii',t}$ | Binary variable; 1 if node $i'$ is connected to node $i$ and energized during time span $t$, 0 otherwise. |

*Parameters*

| | |
|---|---|
| $\Delta t$ | Length of a single time span. |
| $M/\varepsilon$ | A large/small positive number. |
| $T_{j,ii'}$ | Time spans spent traveling from node $i$ to node $i'$ for MER $j$. |
| $W_k$ | Capacity consumed by Mod $k$. |
| $A_j$ | Carrying capacity Carr $j$. |
| $P^{\mathrm{c.S}}_{k,\max}/P^{\mathrm{d.S}}_{k,\max}$ | Maximum charging/discharging power of Mod $k$. |
| $S_{k,\mathrm{Mod}}$ | Rated apparent power of Mod $k$. |
| $E_k$ | Energy capacity of Mod $k$. |
| $e^c_k/e^d_k$ | Charging/discharging efficiency of Mod $k$. |
| $soc_{k,0}$ | Initial SOC of Mod $k$ at the start of scheduling. |
| $SOC_{k,\min}/SOC_{k,\max}$ | Lower/upper bound of allowable range of $SOC_{k,t}$. |
| $P^{\mathrm{G}}_{m,\max}/Q^{\mathrm{G}}_{m,\max}$ | Maximum active/reactive power output of MEG $m$. |
| $S_{m,\mathrm{MEG}}$ | Rated apparent power of MEG $m$. |
| $B_{m,\max}$ | Maximum fuel consumption of MEG $m$ at full load. |
| $p_{m,l}$ | Load level $l$ to depict the fuel consumption characteristic of MEG $m$. $l\in\{1, 2, \ldots, L\}$, where $L$ is the number of those levels. |
| $y_{m,l}$, $z_{m,l}$ | Coefficients of piecewise linearization for fuel consumption characteristic of MEG $m$ within the interval $[p_{m,l-1}, p_{m,l}]$. |
| $F_{m/h/i}$ | Available fuel storage capacity of MEG $m$/FT $h$/node (depot) $i$. |
| $\upsilon_{h,\mathrm{in}}/\upsilon_{h,\mathrm{out}}$ | Maximum rate of fuel input/output of FT $h$. |
| $sof_{i,0}$ | Initial SOF of MEG/FT/node/depot $i$ at the start of scheduling. |
| $i_r$ | The substation node. |
| $P^{\mathrm{L}}_{i,t}/Q^{\mathrm{L}}_{i,t}$ | Active/reactive load at node $i$ during time span $t$. |
| $a_{1,i}$, $a_{2,i}$ | Coefficients used to determine $P^{\mathrm{IN}}_{i,t}/Q^{\mathrm{IN}}_{i,t}$. |
| $r_{i'i}/x_{i'i}$ | Resistance/reactance of branch $(i', i)$. |
| $V_{i,\min}/V_{i,\max}$ | Lower/upper bound of the allowable voltage magnitude at node $i$. |
| $S_{i'i,max}$ | Apparent power capacity of branch $(i', i)$. |
| $w_i$ | Priority weight of the load at node $i$. |
| $\varphi_{travel}/\varphi_{fuel}$ | Cost coefficient for MERs for traveling/fuel exchange. |

# I. Introduction

MOBILE energy resources (MERs) are powerful tools that make distribution systems (DSs) respond to disasters in a rapid and flexible way [1]. The intelligent use of two main types of MERs, *i.e.*, mobile energy storage systems (MESSs) and mobile emergency generators (MEGs), has been highlighted in recent years due to their effectiveness in enhancing power system resilience and economics [2], [3].

To enhance DS resilience, in [4], MEGs are prepositioned prior to a disaster and repositioned to energize the loads after the disaster and resulting damage are known. In [5], the dynamic routing of MERs is further considered. In [6] and [7], MESSs are routed among several parts of a DS or several DSs to realize optimal energy allocation under emergency situations and the DS reconfiguration is involved, which means that the DS topology can be changed to coordinate with the connection of MESSs. In [8], repair crew dispatch is further considered along with the routing and scheduling of MERs and DS reconfiguration for DS restoration. In addition, MESSs also contribute to the improvement in power system economics, whether for bulk power systems [9] or DSs [10]. In [11], the investment decision is studied considering the operation of MESSs during both normal and emergency conditions. From a review of the relevant studies, MERs can truly act as effective "first-aid boxes" for DS service restoration. Compared to the common stationary energy resources, MERs bring higher flexibility to DS restoration, and thus endow DSs with more feasibility and a better response to disasters. Now, standing on the shoulders of the authors of these studies, a conjecture may arise: Can this flexibility be further extended?

To answer this question, we focus our attention on the MESS. Note that in the present studies, the energy storage module (Mod) and carrier (Carr) of an MESS are "fastened" together to be scheduled as one whole, which implies that equivalently only one Mod is contained inside. This restriction creates a gap between the research and the practical fields, if we focus on the state of the art of energy storage and MESS. As a vital characteristic of battery-based energy storage applications and development, the modularity enables energy storage systems to be scalable within a wide range regarding the power or energy capacity [12], [13]. In brief, one can tailor the desired energy storage capacity by easily assembling multiple standard energy storage units, similar to building blocks [14], [15]. In addition, most MESS solutions currently on the market or in projects are containerized and towable [3], *e.g.*, the MESSs provided by RES [16], Consolidated Edison [17], and Aggreko [18]. Thus, these MESSs are inherently separable, *i.e.*, the Mod can be easily loaded on/unloaded from the Carr. Given the above, the idea of a separable MESS (SMESS) is conceived in this paper: We release the above restriction and "split" the only Mod of an MESS into multiple smaller Mods and then let these Mods be independently scattered by the Carr throughout the DS, similar to "one rocket launching multiple satellites". By this action, a more decentralized allocation of energy storage as back-up sources can be realized. Note that the traditional MESS solution is included in the SMESS since we need only to make the Mods and the Carr of an SMESS always be together. Thus, the SMESS solution may be able to endow DS restoration with more flexibility and feasibility, which is attractive as it offers the possibility to push the MERs application to a higher level.

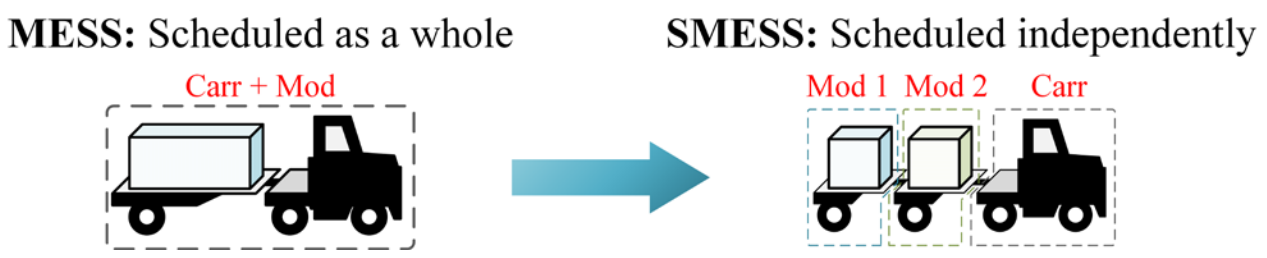

Fig. 1. Evolution from MESS to SMESS.

Then, we move on to MEGs. Most of them are currently powered by fossil fuels. Once an MEG is deployed somewhere in the DS, there should be adequate fuel to ensure its operation for a period of time. Fuel delivery is an inevitable task in deploying MEGs for DS restoration [3], [19] and a critical issue related to supply resilience, as introduced by the EPRI [20]. However, only a few studies involve the scheduling of MEGs, not to mention that almost all of related studies commonly bypassed the fuel delivery issue and left it out of the scheduling by assuming that, *e.g.*, the adequate fuel has been preallocated [5], [8]. Extreme weather striking DSs typically occurs suddenly and is hard to anticipate, so it is likely that such preallocation of fuel may not be fully accomplished and that this assumption is too ideal. In many circumstances, it is necessary to schedule fuel tankers (FTs) jointly with MEGs in a coordinated way to guarantee adequate fuel for MEGs operation. To the best of our knowledge, such joint scheduling has not yet been developed.

To bridge the above gaps, in this paper, we propose a method for jointly scheduling SMESSs, MEGs, and FTs, along with the dynamic DS reconfiguration, for DS restoration. The main contributions of this paper are threefold:

1) A novel SMESS concept is proposed, and the constraints for scheduling SMESSs are derived. In contrast to a conventional MESS solution, an SMESS can carry multiple Mods and scatter them independently throughout the DS. 2) Fuel delivery coordinated with the scheduling of MEGs is considered, and the constraints for scheduling FTs are derived. 3) A joint scheduling model for DS restoration, involving the routing and scheduling of SMESSs, MEGs, and FTs and the scheduling of the dynamic DS reconfiguration, is developed in the form of mixed-integer linear programming (MILP) and its advantage in boosting the DS resilience is verified by tests.

The rest of this paper is organized as follows: Section II derives the constraints for routing and scheduling of SMESSs, MEGs and FTs; Section III gives the constraints for dynamic DS reconfiguration and operation; Section IV summarizes the joint scheduling model for DS restoration; Section V provides the numerical studies; and Section VI concludes this paper.

## II. Scheduling of SMESSs, MEGs and Fuel Delivery

### A. *Evolution from MESS to SMESS*

According to the investigation of [3], most of the MESS products at present are towable, *i.e.*, they are combinations of tractors and trailers. As depicted in Fig .1, the key evolutions from the scheduling of MESSs to SMESSs can be recognized as follows: 1) A Carr (*i.e.*, a tractor) carries multiple detachable Mods rather than one attached to it. Specifically, each Mod is mounted onto a trailer, and some Mods are then towed in series by the Carr. 2) The Carr and each of the carried Mods are treated as independent rather than integrated components in scheduling. Once a Carr arrives somewhere in a DS, it can drop one or more of the Mods it is carrying; and when a Carr is about to depart, the Carr can tow away one or more of the Mods located here.

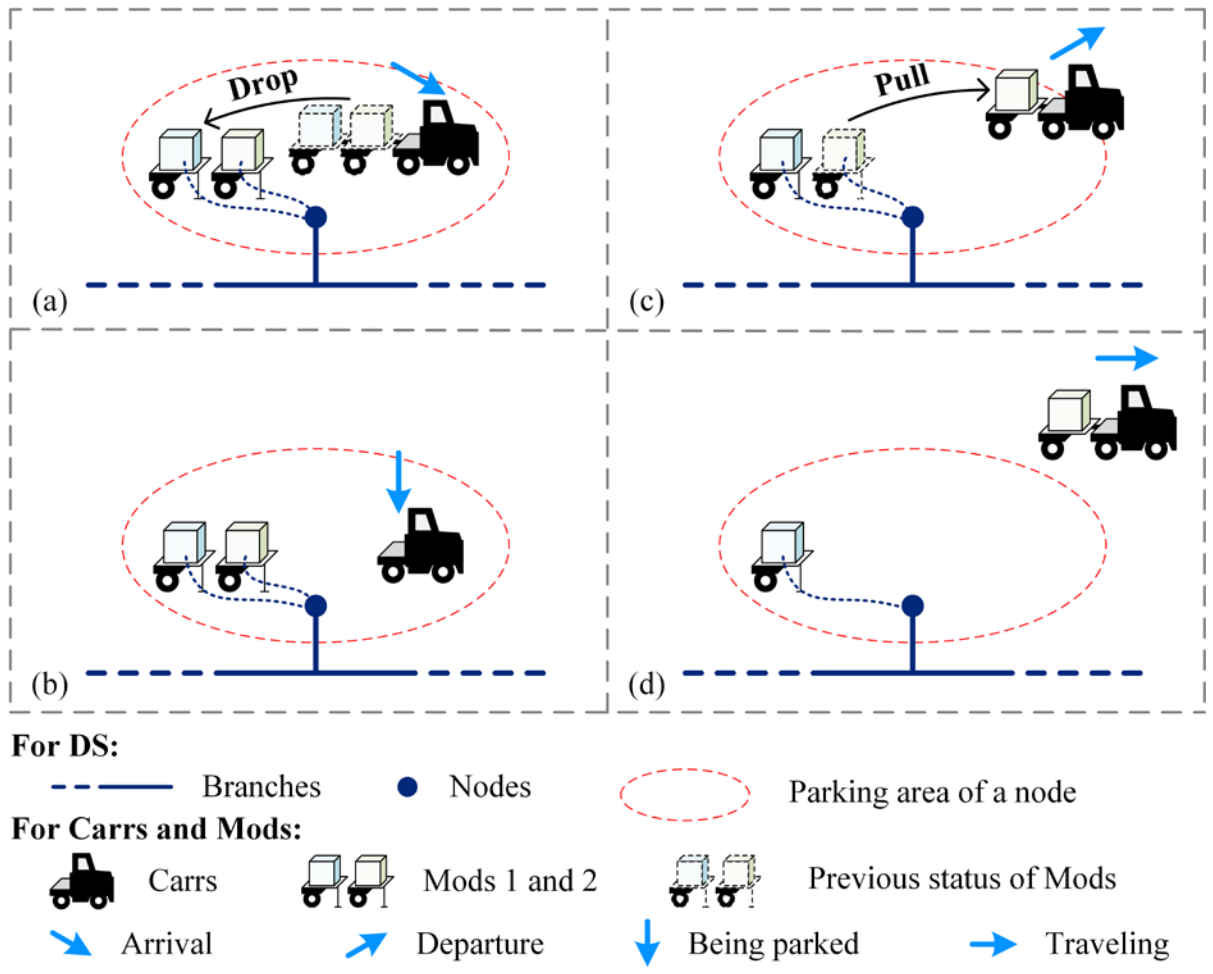

Fig. 2. Illustration of SMESS behavior. (a) A Carr arrives at a node; (b) a Carr is being parked at a node; (c) a Carr departs from a node; and (d) a Carr is traveling without contacting to any node.

### B. *Routing of MERs*

In our prior work [21], we proposed a novel mobility model for the routing of MERs, which was shown to have a good computational efficiency. This model, as reviewed in (1), is used for the routing of all three kinds of MERs, *i.e.*, SMESSs (actually the Carrs of SMESSs), MEGs and FTs in this paper.

$$\sum_{i\in\mathcal{N}_{\mathrm{M}}} x_{j,i,t} + \sum_{i\in\mathcal{N}_{\mathrm{M}}} v_{j,i,t} = 1 \quad, \forall t\in\mathcal{T}\cup\{0\},\ j\in\mathcal{M} \quad (1a)$$

$$\begin{cases} x_{j,i,t+1} \ge x_{j,i,t} + 1.2\left(v_{j,i,t} - v_{j,i,t+1}\right) + 0.4\left(\sum_{i\in\mathcal{N}_{\mathrm{M}}} v_{j,i,t} - \sum_{i\in\mathcal{N}_{\mathrm{M}}} v_{j,i,t+1}\right) - 0.8 \\ x_{j,i,t+1} \le x_{j,i,t} + \left(v_{j,i,t} - v_{j,i,t+1}\right) - 0.5\left(\sum_{i\in\mathcal{N}_{\mathrm{M}}} v_{j,i,t} - \sum_{i\in\mathcal{N}_{\mathrm{M}}} v_{j,i,t+1}\right) + 0.7 \end{cases} \quad (1b)$$

$$, \forall t\in\{\mathcal{T}\cup\{0\}\}\setminus\{D\},\ j\in\mathcal{M}, i\in\mathcal{N}_{\mathrm{M}}$$

$$\begin{cases} S_{j,t} \ge x_{j,i,t-1}\sum_{i'\in\mathcal{N}_{\mathrm{M}}} T_{j,ii'} + \sum_{i'\in\mathcal{N}_{\mathrm{M}}}\left(v_{j,i',t} T_{j,ii'}\right) - \sum_{i'\in\mathcal{N}_{\mathrm{M}}} T_{j,ii'} \quad, \forall i\in\mathcal{N}_{\mathrm{M}} \\ S_{j,t} \ge 0 \end{cases} \quad (1c)$$

$$, \forall t\in\mathcal{T},\ j\in\mathcal{M}$$

$$R_{j,t} = R_{j,t-1} + S_{j,t} - \sum_{i\in\mathcal{N}_{\mathrm{M}}} v_{j,i,t-1} \quad, \forall t\in\mathcal{T},\ j\in\mathcal{M} \quad (1d)$$

$$R_{j,t}/M \le \sum_{i\in\mathcal{N}_{\mathrm{M}}} v_{j,i,t} \le R_{j,t} \quad, \forall\ t\in\mathcal{T}\cup\{0\},\ j\in\mathcal{M} \quad (1e)$$

$$\begin{cases} \omega_{j,t} \ge \sum_{i\in\mathcal{N}_{\mathrm{M}}} v_{j,i,t-1} + \sum_{i\in\mathcal{N}_{\mathrm{M}}} v_{j,i,t} - 2 + \varepsilon \\ -\left(1-\omega_{j,t}\right) \le v_{j,i,t} - v_{j,i,t-1} \le \left(1-\omega_{j,t}\right),\ \forall i\in\mathcal{N}_{\mathrm{M}} \end{cases} \quad (1f)$$

$$, \forall t\in\mathcal{T},\ j\in\mathcal{M}$$

$$x_{j,i_j,0} = 1,\ S_{j,0} = 0,\ R_{j,0} = 0,\ \omega_{j,0} = 0 \quad, \forall j\in\mathcal{M} \quad (1g)$$

Specifically, constraint (1a) indicates that, for any MER, there is only one state during any time span, *i.e.*, it can only be parked at or travel to somewhere. Constraint (1b) represents the state transition between parking state and traveling state of an MER. Constraint (1c) recognizes the necessary travel time required for a travel of an MER. By (1c), when a travel from node $i_1$ to node $i_2$ starts (for MER $j$), $S_{j,t}$ is forces to be the travel time required, *i.e.*, $T_{j,i_1i_2}$. Otherwise, when no travel starts, $S_{j,t}$ is forced to be 0. Then, constraint (1d) defines the residual travel time required to be consumed by an MER. When a travel starts, $R_{j,t}$ is equal to $S_{j,t}$; and as $R_{j,t}$ decreases, the travel is gradually coming to the end. Constraint (1e) restricts the traveling state of an MER based on the value of $R_{j,t}$. Constraint (1f) maintains the direction during a travel. Constraints (1g) denotes the initial condition for the routing. The detailed explanation and derivation of (1a)-(1g), as well as the comparisons between this mobility model and the other existing mobility models, can be found in our prior work [21].

The $\mathcal{M}$ and $\mathcal{N}_{\mathrm{M}}$ in (1a)-(1g) represent the MERs and the nodes that support their routing, respectively, and we write them in this way just for convenience. Specifically, the above model can realize the routing of MEGs (or FTs) by replacing $\mathcal{M}$ with $\mathcal{M}_{\mathrm{G}}$ (or $\mathcal{M}_{\mathrm{F}}$) and $\mathcal{N}_{\mathrm{M}}$ with $\mathcal{N}_{\mathrm{G}}\cup\mathcal{N}_{\mathrm{DP}}$, and realizes the routing of SMESSs by replacing $\mathcal{M}$ with $\mathcal{M}_{\mathrm{S}}$ and $\mathcal{N}_{\mathrm{M}}$ with $\mathcal{N}_{\mathrm{S}}$.

### C. Scheduling of SMESSs

A Mod always belongs to either a node or a Carr, and the carrying capacity of Carrs cannot be exceeded. Thus, we have

$$\sum\nolimits_{i\in\mathcal{N}_{\mathrm{S}}}\zeta_{k,i,t}+\sum\nolimits_{j\in\mathcal{M}_{\mathrm{S}}}\gamma_{k,j,t}=1 \quad ,\forall t\in\mathcal{T}\cup\{0\},k\in\mathcal{K} \quad (2a)$$

$$\sum\nolimits_{k\in\mathcal{K}}W_k\gamma_{k,j,t}\le A_j \quad ,\forall t\in\mathcal{T},j\in\mathcal{M}_{\mathrm{S}} \quad (2b)$$

$$\zeta_{k,i_k,0}=1 \quad ,\forall k\in\mathcal{K} \quad (2c)$$

In addition, Mod $k$ is initially located at node $i_k$, as denoted by (2c). All the possible interactive behaviors in the SMESS itself and between the SMESS and DS are enumerated in Fig. 2, upon which the scheduling model of the SMESS can be derived from the perspectives of both Carrs and nodes.

#### 1) Between Carrs and Mods

*Scenario 1 for Carrs*. It is reasonable to assume that a node always dominate all the Mods located at it, as shown in Fig.2 (a)-(b). In other words, for a Carr, it does not own any Mod when it arrives or is parked at a node. Thus, we have

$$\gamma_{k,j,t}\le 1-\sum\nolimits_{i\in\mathcal{N}_{\mathrm{S}}}x_{j,i,t} \quad ,\forall t\in\mathcal{T},j\in\mathcal{M}_{\mathrm{S}},k\in\mathcal{K} \quad (3a)$$

*Scenario 2 for Carrs*. Then, when a Carr departs from a node, it can carry away some of the Mods owned by this node, as shown in Fig. 2 (c). We can formulate that as

$$\gamma_{k,j,t}-\zeta_{k,i,t-1}\le x_{j,i,t}+1-x_{j,i,t-1} \quad ,\forall t\in\mathcal{T},j\in\mathcal{M}_{\mathrm{S}},i\in\mathcal{N}_{\mathrm{S}},k\in\mathcal{K} \quad (3b)$$

*Scenario 3 for Carrs*. Other than the above scenarios, *i.e.*, when a Carr is traveling, as shown in Fig. 2 (d), the Carr should hold the Mods on it. This can be formulated as

$$-\left(\sum\nolimits_{i\in\mathcal{N}_{\mathrm{S}}}x_{j,i,t-1}+\sum\nolimits_{i\in\mathcal{N}_{\mathrm{S}}}x_{j,i,t}\right)\le\gamma_{k,j,t}-\gamma_{k,j,t-1}\le \sum\nolimits_{i\in\mathcal{N}_{\mathrm{S}}}x_{j,i,t-1}+\sum\nolimits_{i\in\mathcal{N}_{\mathrm{S}}}x_{j,i,t} \quad ,\forall t\in\mathcal{T},j\in\mathcal{M}_{\mathrm{S}},k\in\mathcal{K} \quad (3c)$$

#### 2) Between Nodes and Mods

*Scenario 1 for nodes*. If a Carr carrying a Mod arrives at a node, then this node must own this Mod, as shown in Fig. 2 (a).

*Scenario 2 for nodes*. Otherwise, $\zeta_{k,i,t}\le\zeta_{k,i,t-1}$ should be ensured, which means that node $i$ cannot obtain Mod $k$ when $\zeta_{k,i,t-1}=0$ or may lose Mod $k$ to some Carr when $\zeta_{k,i,t-1}=1$.

To distinguish the above two scenarios, a binary variable $\alpha_{j,i,k,t}$ is defined to indicate whether Carr $j$ carrying Mod $k$ arrives at node $i$ during time span $t$, *i.e.*, $\alpha_{j,i,k,t}=(\neg x_{j,i,t-1})\wedge x_{j,i,t}\wedge\gamma_{k,j,t-1}$. According to [22] and [23], we can formulate it by

$$\begin{cases}\alpha_{j,i,k,t}\le 1-x_{j,i,t-1} \ ; \ \alpha_{j,i,k,t}\le x_{j,i,t} \ ; \\ \alpha_{j,i,k,t}\le\gamma_{k,j,t-1} \ ; \ \alpha_{j,i,k,t}\ge -x_{j,i,t-1}+x_{j,i,t}+\gamma_{k,j,t-1}-1\end{cases} \quad (4a)$$
$$,\forall t\in\mathcal{T},i\in\mathcal{N}_{\mathrm{S}},j\in\mathcal{M}_{\mathrm{S}},k\in\mathcal{K}$$

Thus, we have

$$\zeta_{k,i,t}\ge\sum\nolimits_{j\in\mathcal{M}_{\mathrm{S}}}\alpha_{j,i,k,t} \quad ,\forall t\in\mathcal{T},i\in\mathcal{N}_{\mathrm{S}},k\in\mathcal{K} \quad (4b)$$

$$\zeta_{k,i,t}-\zeta_{k,i,t-1}\le\sum\nolimits_{j\in\mathcal{M}_{\mathrm{S}}}\alpha_{j,i,k,t} \quad ,\forall t\in\mathcal{T},i\in\mathcal{N}_{\mathrm{S}},k\in\mathcal{K} \quad (4c)$$

Constraints (4b) and (4c) represent *Scenario 1 for nodes* and *Scenario 2 for nodes*, respectively.

*Scenario 3 for nodes*. In addition, if node $i$ is "undisturbed", *i.e.*, no Carr arrives at or departs from it, as shown in Fig. 2 (d), each Mod should maintain the status regarding it, *i.e.*, $\zeta_{k,i,t}=\zeta_{k,i,t-1}$. This can be naturally realized by the above constraints.

Suppose that node $i$ is undisturbed during time span $t$, and we know from (4a) that $\Sigma_{j\in\mathcal{M}_{\mathrm{S}}}\alpha_{j,i,k,t}=0$. When $\zeta_{k,i,t-1}=0$, then $\zeta_{k,i,t}=0$ from (4c). On the other hand, when $\zeta_{k,i,t-1}=1$, let us consider where this Mod $k$ can go during the next time span $t$: *i)* For any other node $i_2$ ($i_2\neq i$), we know from (2a) that $\gamma_{k,j,t-1}=0$ for any Carr $j$ and $\zeta_{k,i_2,t-1}=0$; then, based on (4a) and (4c), we can obtain $\Sigma_{j\in\mathcal{M}_{\mathrm{S}}}\alpha_{j,i_2,k,t}=0$ and $\zeta_{k,i_2,t}=0$. *ii)* For any Carr $j$, if it just departed from some other node $i_2$ ($i_2\neq i$), *i.e.*, $x_{j,i_2,t-1}=1$ and $x_{j,i_2,t}=0$, we know that $\zeta_{k,i_2,t-1}=0$, and then, based on (3b), $\gamma_{k,j,t}=0$; if Carr $j$ is traveling during time spans $t-1$ and $t$, it is clear from (3c) that $\gamma_{k,j,t-1}=0$ and $\gamma_{k,j,t}=0$; in addition, if Carr $j$ is parked at some node, there clearly is $\gamma_{k,j,t}=0$ due to (3a). From *i)* and *ii)*, Mod $k$ at node $i$ cannot go anywhere but instead stay only at node $i$. Thus, no additional constraints are needed for *Scenario 3 for nodes*.

#### 3) Operation of Mods

$$c_{k,i,t}+d_{k,i,t}\le\zeta_{k,i,t} \quad ,\forall t\in\mathcal{T},k\in\mathcal{K},i\in\mathcal{N}_{\mathrm{S}} \quad (5a)$$

$$\begin{cases}0\le P^{\mathrm{c.S}}_{k,i,t}\le c_{k,i,t}P^{\mathrm{c.S}}_{k,\max}\\ 0\le P^{\mathrm{d.S}}_{k,i,t}\le d_{k,i,t}P^{\mathrm{d.S}}_{k,\max}\\ -S_{k,\mathrm{Mod}}\zeta_{k,i,t}\le Q^{\mathrm{S}}_{k,i,t}\le S_{k,\mathrm{Mod}}\zeta_{k,i,t}\end{cases} \quad ,\forall t\in\mathcal{T},k\in\mathcal{K},i\in\mathcal{N}_{\mathrm{S}} \quad (5b)$$

$$\left[\sum\nolimits_{i\in\mathcal{N}_{\mathrm{S}}}\left(P^{\mathrm{d.S}}_{k,i,t}-P^{\mathrm{c.S}}_{k,i,t}\right)\right]^2+\left(\sum\nolimits_{i\in\mathcal{N}_{\mathrm{S}}}Q^{\mathrm{S}}_{k,i,t}\right)^2\le S^2_{k,\mathrm{Mod}} \quad ,\forall t\in\mathcal{T},k\in\mathcal{K} \quad (5c)$$

$$SOC_{k,t}=SOC_{k,t-1}+\left(e^{\mathrm{c}}_k\sum\nolimits_{i\in\mathcal{N}_{\mathrm{S}}}P^{\mathrm{c.S}}_{k,i,t}-\sum\nolimits_{i\in\mathcal{N}_{\mathrm{S}}}P^{\mathrm{d.S}}_{k,i,t}\Big/e^{\mathrm{d}}_k\right)\Delta t\Big/E_k \quad ,\forall t\in\mathcal{T},k\in\mathcal{K} \quad (5d)$$

$$SOC_{k,0}=soc_{k,0} \quad ,\forall k\in\mathcal{K} \quad (5e)$$

$$SOC_{k,\min}\le SOC_{k,t}\le SOC_{k,\max} \quad ,\forall t\in\mathcal{T},k\in\mathcal{K} \quad (5f)$$

Constraint (5a) restricts the charging/discharging mode of Mods. Constraints (5b)-(5c) bound the power output, and the reactive power support of Mods is considered, as in [10]. Constraint (5d) represents the relationship of the state of charge (SOC) of a Mod between two adjacent time spans. Constraint (5e) indicates the initial SOC of the Mods. Constraint (5f) restricts the allowable range of SOC of the Mods.

### D. Operation of MEGs

$$\begin{cases} 0 \le P_{m,i,t}^{\mathrm{G}} \le x_{m,i,t} P_{m,\max}^{\mathrm{G}} \\ 0 \le Q_{m,i,t}^{\mathrm{G}} \le x_{m,i,t} Q_{m,\max}^{\mathrm{G}} \end{cases} ,\forall t \in \mathcal{T}, m \in \mathcal{M}_{\mathrm{G}}, i \in \mathcal{N}_{\mathrm{G}} \quad (6a)$$

$$\left(\sum\nolimits_{i\in\mathcal{N}_{\mathrm{G}}} P_{m,i,t}^{\mathrm{G}}\right)^2 + \left(\sum\nolimits_{i\in\mathcal{N}_{\mathrm{G}}} Q_{m,i,t}^{\mathrm{G}}\right)^2 \le S_{m,\mathrm{MEG}}^2 ,\forall t \in \mathcal{T}, m \in \mathcal{M}_{\mathrm{G}} \quad (6b)$$

Constraints (6a)-(6b) bound the power output of MEGs.

### E. Scheduling of FTs

The gross fuel demand of an MEG at a node is restricted as

$$0 \le B_{m,i,t} \le x_{m,i,t} B_{m,\max} \quad ,\forall t \in \mathcal{T}, m \in \mathcal{M}_{\mathrm{G}}, i \in \mathcal{N}_{\mathrm{G}} \quad (7a)$$

$$B_{m,i,t} = 0 \quad ,\forall t \in \mathcal{T}, m \in \mathcal{M}_{\mathrm{G}}, i \in \mathcal{N}_{\mathrm{DP}} \quad (7b)$$

Manufacturers commonly provide the fuel consumption rate of MEGs at several load levels, *e.g.*, 1/4, 1/2, 3/4, and full load, as in [24]. Hence, we can formulate the gross fuel demand of an MEG as the following piecewise linearization:

$$\sum\nolimits_{i\in\mathcal{N}_{\mathrm{G}}} B_{m,i,t} = y_{m,l} \sum\nolimits_{i\in\mathcal{N}_{\mathrm{G}}} P_{m,i,t}^{\mathrm{G}} + z_{m,l} \quad ,\text{if} \sum\nolimits_{i\in\mathcal{N}_{\mathrm{G}}} P_{m,i,t}^{\mathrm{G}} \in \left[p_{m,l-1}, p_{m,l}\right] \quad (7c)$$

Equation (7c) can be further formulated as follows [22], [25]:

$$\begin{cases} -M\left(1-\tau_{m,t,l}\right) \le \sum\nolimits_{i\in\mathcal{N}_{\mathrm{G}}} B_{m,i,t} - \left(y_{m,l}\sum\nolimits_{i\in\mathcal{N}_{\mathrm{G}}} P_{m,i,t}^{\mathrm{G}} + z_{m,l}\right) \\ \sum\nolimits_{i\in\mathcal{N}_{\mathrm{G}}} B_{m,i,t} - \left(y_{m,l}\sum\nolimits_{i\in\mathcal{N}_{\mathrm{G}}} P_{m,i,t}^{\mathrm{G}} + z_{m,l}\right) \le M\left(1-\tau_{m,t,l}\right) \\ p_{m,l-1} - \sum\nolimits_{i\in\mathcal{N}_{\mathrm{G}}} P_{m,i,t}^{\mathrm{G}} \le M\left(1-\tau_{m,t,l}\right) \\ \sum\nolimits_{i\in\mathcal{N}_{\mathrm{G}}} P_{m,i,t}^{\mathrm{G}} - p_{m,l} \le M\left(1-\tau_{m,t,l}\right) \end{cases} \quad (7d)$$
$$,\forall t \in \mathcal{T}, m \in \mathcal{M}_{\mathrm{G}}, l \in \{1,3,\dots,L\}$$

$$\sum\nolimits_{l\in\{1,3,\dots,L\}} \tau_{m,t,l} = 1 \quad ,\forall t \in \mathcal{T}, m \in \mathcal{M}_{\mathrm{G}} \quad (7e)$$

Similar to the SOC for Mods, we introduce the "state of fuel (SOF)" for FTs, MEGs, nodes, and fuel depots to represent the ratio of fuel currently stored in them to their fuel capacities. The nodes that support the MEGs connection are considered to have the capability of fuel storage, *i.e.*, fuel can be stored at the locations of these nodes (*e.g.*, by means of fuel tanks there). In addition, even if some of the nodes cannot store fuel, we can just set their fuel capacity $F_i$ in (7i) to a positive number near to 0. Thus, the fuel supplemented from FTs and the fuel consumed by MEGs at these nodes are balanced in each time span under constraints (7i) and (7o).

The fuel carried by an MEG may not always be enough to feed itself and the extra fuel demanded can be derived as "if gross fuel demand $\Sigma_{i\in\mathcal{N}_{\mathrm{G}}} B_{m,i,t}$ is less than the fuel on MEG $m$ itself $SOF_{m,t-1}\cdot F_m$, then MEG $m$ is self-sufficient during time span $t$ and no extra fuel is needed; otherwise, the extra fuel $\Sigma_{i\in\mathcal{N}_{\mathrm{G}}} B_{m,i,t} - SOF_{m,t-1}\cdot F_m$ is needed." We can introduce a binary variable $b_{m,t}$ as the indicator of $\Sigma_{i\in\mathcal{N}_{\mathrm{G}}} B_{m,i,t} < SOF_{m,t-1}\cdot F_m$, and the above logic is formulated as

$$\begin{cases} 0 \le B_{m,i,t}^{+} \le B_{m,i,t} \ ,\forall i \in \mathcal{N}_{\mathrm{G}} \cup \mathcal{N}_{\mathrm{DP}} \\ -F_m b_{m,t} \le \sum\nolimits_{i\in\mathcal{N}_{\mathrm{G}}} B_{m,i,t} - SOF_{m,t-1}F_m \le B_{m,\max}\left(1-b_{m,t}\right) \\ \sum\nolimits_{i\in\mathcal{N}_{\mathrm{G}}} B_{m,i,t}^{+} \le B_{m,\max}\left(1-b_{m,t}\right) \\ 0 \le \sum\nolimits_{i\in\mathcal{N}_{\mathrm{G}}} B_{m,i,t}^{+} - \left(\sum\nolimits_{i\in\mathcal{N}_{\mathrm{G}}} B_{m,i,t} - SOF_{m,t-1}F_m\right) \le F_m b_{m,t} \end{cases} \quad (7f)$$
$$,\forall t \in \mathcal{T}, m \in \mathcal{M}_{\mathrm{G}}$$

The SOF of relevant facilities are restricted by (7g)-(7n).

$$SOF_{m,t} = SOF_{m,t-1} - \sum\nolimits_{i\in\mathcal{N}_{\mathrm{G}}\cup\mathcal{N}_{\mathrm{DP}}} \left(B_{m,i,t} - B_{m,i,t}^{+} - G_{m,i,t}\right) \Big/ F_m \quad ,\forall t \in \mathcal{T}, m \in \mathcal{M}_{\mathrm{G}} \quad (7g)$$

$$SOF_{h,t} = SOF_{h,t-1} - \sum\nolimits_{i\in\mathcal{N}_{\mathrm{G}}\cup\mathcal{N}_{\mathrm{DP}}} D_{h,i,t} \Big/ F_h \quad ,\forall t \in \mathcal{T}, h \in \mathcal{M}_{\mathrm{F}} \quad (7h)$$

$$SOF_{i,t} = SOF_{i,t-1} + \left[\sum\nolimits_{h\in\mathcal{M}_{\mathrm{F}}} D_{h,i,t} - \sum\nolimits_{m\in\mathcal{M}_{\mathrm{G}}} \left(B_{m,i,t}^{+} + G_{m,i,t}\right)\right] \Big/ F_i \quad ,\forall t \in \mathcal{T}, i \in \mathcal{N}_{\mathrm{G}} \cup \mathcal{N}_{\mathrm{DP}} \quad (7i)$$

$$\iota_{m,i,t} \le x_{m,i,t} \quad ,\forall t \in \mathcal{T}, m \in \mathcal{M}_{\mathrm{G}}, i \in \mathcal{N}_{\mathrm{G}} \cup \mathcal{N}_{\mathrm{DP}} \quad (7j)$$

$$-F_m \iota_{m,i,t} \le G_{m,i,t} \le F_m \iota_{m,i,t} \ ,\forall t \in \mathcal{T}, m \in \mathcal{M}_{\mathrm{G}}, i \in \mathcal{N}_{\mathrm{G}} \cup \mathcal{N}_{\mathrm{DP}} \quad (7k)$$

$$\iota_{h,i,t} \le x_{h,i,t} \quad ,\forall t \in \mathcal{T}, h \in \mathcal{M}_{\mathrm{F}}, i \in \mathcal{N}_{\mathrm{G}} \cup \mathcal{N}_{\mathrm{DP}} \quad (7l)$$

$$-\left(\upsilon_{h,\mathrm{in}}\Delta t\right)\iota_{h,i,t} \le D_{h,i,t} \le \left(\upsilon_{h,\mathrm{out}}\Delta t\right)\iota_{h,i,t} \quad ,\forall t \in \mathcal{T}, h \in \mathcal{M}_{\mathrm{F}}, i \in \mathcal{N}_{\mathrm{G}} \cup \mathcal{N}_{\mathrm{DP}} \quad (7m)$$

$$SOF_{i,0} = sof_{i,0} \quad ,\ \forall i \in \mathcal{M}_{\mathrm{G}} \cup \mathcal{M}_{\mathrm{F}} \cup \mathcal{N}_{\mathrm{G}} \cup \mathcal{N}_{\mathrm{DP}} \quad (7n)$$

$$0 \le SOF_{i,t} \le 1 \quad ,\forall t \in \mathcal{T}, i \in \mathcal{M}_{\mathrm{G}} \cup \mathcal{M}_{\mathrm{F}} \cup \mathcal{N}_{\mathrm{G}} \cup \mathcal{N}_{\mathrm{DP}} \quad (7o)$$

## III. Dynamic DS Reconfiguration and Operation

### A. Dynamic DS Reconfiguration

We use the flow-based model presented by [26] to formulate the constraints for dynamic DS reconfiguration as follows:

$$\sum\nolimits_{(i',i_r)\in\mathcal{L}} f_{i'i_r,t}^{i_x} - \sum\nolimits_{(i_r,i')\in\mathcal{L}} f_{i_r i',t}^{i_x} = -1 \quad ,\forall t \in \mathcal{T}, i_x \in \mathcal{N} \setminus i_r \quad (8a)$$

$$\sum\nolimits_{(i',i_x)\in\mathcal{L}} f_{i'i_x,t}^{i_x} - \sum\nolimits_{(i_x,i')\in\mathcal{L}} f_{i_x i',t}^{i_x} = 1 \quad ,\forall t \in \mathcal{T}, i_x \in \mathcal{N} \setminus i_r \quad (8b)$$

$$\sum\nolimits_{(i',i)\in\mathcal{L}} f_{i'i,t}^{i_x} - \sum\nolimits_{(i,i')\in\mathcal{L}} f_{ii',t}^{i_x} = 0 \quad ,\forall t \in \mathcal{T}, i_x \in \mathcal{N} \setminus i_r \ , i \in \mathcal{N} \setminus \{i_x, i_r\} \quad (8c)$$

$$0 \le f_{i'i,t}^{i_x} \le \lambda_{i'i,t} \ , 0 \le f_{ii',t}^{i_x} \le \lambda_{ii',t} \quad ,\forall t \in \mathcal{T}\ , i_x \in \mathcal{N} \setminus i_r, (i',i) \in \mathcal{L} \quad (8d)$$

$$\sum\nolimits_{(i',i)\in\mathcal{L}} \left(\lambda_{i'i,t} + \lambda_{ii',t}\right) = |\mathcal{N}| - 1 \quad ,\forall t \in \mathcal{T} \quad (8e)$$

$$\lambda_{i'i,t} + \lambda_{ii',t} = \mu_{i'i,t} \quad ,\forall t \in \mathcal{T}, (i',i) \in \mathcal{L} \quad (8f)$$

$$\kappa_{i'i,t} \le \mu_{i'i,t} \quad ,\forall t \in \mathcal{T}, (i',i) \in \mathcal{L} \quad (8g)$$

Constraints (8a)-(8g) always ensure the radiality of the DS, and specific explanation can be found in [26]. The above model ensures the radiality in dynamic DS reconfiguration by a two-step way based on graph theory. First, all the fictitious spanning trees of the DS are represented by (8a)-(8f). In contrast to "actual", the word "fictitious" means that damages or faults that exist in the actual DS are not considered here. Thus, a fictitious DS is the same as the actual DS but without any damage [8]. The fictitious spanning trees are spanning trees of the fictitious DS. Next, a feasible topology of the actual DS that can be formed via reconfiguration can be exactly represented by a subset of one fictitious spanning tree, as restricted by (8g). Thus, the actual topology of the DS is determined by "$\kappa_{ii',t}$" rather than "$\mu_{ii',t}$". In addition, the faults in the actual DS are considered by (9i), which also restricts the feasible topology of the actual DS along with (8g).

### *B. Operation of the DS*

The constraints for DS operation are given as (9a)-(9m), based upon the linearized DistFlow model [8], [26], [27].

$$P_{i,t}^{\mathrm{IN}}=a_{1,i}\sum\nolimits_{k\in\mathcal{K}}\left(P_{k,i,t}^{\mathrm{d.S}}-P_{k,i,t}^{\mathrm{c.S}}\right)+a_{2,i}\sum\nolimits_{m\in\mathcal{M}_{\mathrm{G}}}P_{m,i,t}^{\mathrm{G}} \quad ,\forall t\in\mathcal{T},i\in\mathcal{N} \tag{9a}$$

$$Q_{i,t}^{\mathrm{IN}}=a_{1,i}\sum\nolimits_{k\in\mathcal{K}}Q_{k,i,t}^{\mathrm{S}}+a_{2,i}\sum\nolimits_{m\in\mathcal{M}_{\mathrm{G}}}Q_{m,i,t}^{\mathrm{G}} \quad ,\forall t\in\mathcal{T},i\in\mathcal{N} \tag{9b}$$

$$\sum\nolimits_{(i',i)\in\mathcal{L}}P_{i'i,t}+P_{i,t}^{\mathrm{IN}}-\delta_{i,t}P_{i,t}^{\mathrm{L}}=\sum\nolimits_{(i,i')\in\mathcal{L}}P_{ii',t}\ ,\forall t\in\mathcal{T},i\in\mathcal{N} \tag{9c}$$

$$\sum\nolimits_{(i',i)\in\mathcal{L}}Q_{i'i,t}+Q_{i,t}^{\mathrm{IN}}-\delta_{i,t}Q_{i,t}^{\mathrm{L}}=\sum\nolimits_{(i,i')\in\mathcal{L}}Q_{ii',t}\ ,\forall t\in\mathcal{T},i\in\mathcal{N} \tag{9d}$$

$$\delta_{i,t}\geq\delta_{i,t-1}\quad ,\forall t\in\mathcal{T},i\in\mathcal{N} \tag{9e}$$

$$\begin{cases}V_{i,t}^2\geq V_{i',t}^2-2\left(P_{i'i,t}r_{i'i}+Q_{i'i,t}x_{i'i}\right)-M\left(1-\kappa_{i'i,t}\right)\\ V_{i,t}^2\leq V_{i',t}^2-2\left(P_{i'i,t}r_{i'i}+Q_{i'i,t}x_{i'i}\right)+M\left(1-\kappa_{i'i,t}\right)\end{cases} \quad ,\forall t\in\mathcal{T},\left(i',i\right)\in\mathcal{L} \tag{9f}$$

$$V_{i,\min}^2\leq V_{i,t}^2\leq V_{i,\max}^2\quad ,\forall t\in\mathcal{T},i\in\mathcal{N} \tag{9g}$$

$$P_{i'i,t}^2+Q_{i'i,t}^2\leq\kappa_{i'i,t}S_{i'i,\max}^2\quad ,\forall t\in\mathcal{T},\left(i',i\right)\in\mathcal{L} \tag{9h}$$

$$\begin{cases}\kappa_{i_1i_2,t}=0,\forall\left(i_1,i_2\right)\in\mathcal{L}_{\mathrm{FO},t}\ ;\ \kappa_{i_1i_2,t}=1,\forall\left(i_1,i_2\right)\in\mathcal{L}_{\mathrm{FC},t}\ ;\\ \delta_{i,t}=0\ ,\forall i\in\mathcal{N}_{\mathrm{FO},t}\ ;\ \delta_{i,t}\geq\eta_{i,t}\ ,\forall i\in\mathcal{N}_{\mathrm{FC},t}\end{cases} \quad ,\forall t\in\mathcal{T} \tag{9i}$$

$$\eta_{i_r,t}=1\quad ,\forall t\in\mathcal{T} \tag{9j}$$

$$\begin{cases}\rho_{i,t}\geq\left(a_{1,i}\sum\nolimits_{k\in\mathcal{K}}\zeta_{k,i,t}+a_{2,i}\sum\nolimits_{m\in\mathcal{M}_{\mathrm{G}}}x_{m,i,t}\right)\Big/\left(a_{1,i}\left|\mathcal{K}\right|+a_{2,i}\left|\mathcal{M}_{\mathrm{G}}\right|+1\right)\\ \rho_{i,t}\leq a_{1,i}\sum\nolimits_{k\in\mathcal{K}}\zeta_{k,i,t}+a_{2,i}\sum\nolimits_{m\in\mathcal{M}_{\mathrm{G}}}x_{m,i,t}\end{cases} \quad ,\forall t\in\mathcal{T},i\in\mathcal{N}\setminus\left\{i_r\right\} \tag{9k}$$

$$\begin{cases}\sigma_{i,t}\geq\left(\sum\nolimits_{(i,i')\in\mathcal{L}}\eta_{i',t}\kappa_{ii',t}+\sum\nolimits_{(i',i)\in\mathcal{L}}\eta_{i',t}\kappa_{i'i,t}\right)\Big/M\\ \sigma_{i,t}\leq\sum\nolimits_{(i,i')\in\mathcal{L}}\eta_{i',t}\kappa_{ii',t}+\sum\nolimits_{(i',i)\in\mathcal{L}}\eta_{i',t}\kappa_{i'i,t}\end{cases} \quad ,\forall t\in\mathcal{T},i\in\mathcal{N}\setminus\left\{i_r\right\} \tag{9l}$$

$$\eta_{i,t}\geq\rho_{i,t}\ ,\eta_{i,t}\geq\sigma_{i,t}\ ,\eta_{i,t}\leq\rho_{i,t}+\sigma_{i,t}\quad ,\forall t\in\mathcal{T},i\in\mathcal{N}\setminus\left\{i_r\right\} \tag{9m}$$

The power input at each node is expressed in a general form by (9a)-(9b). The values of $a_{1,i,t}$ and $a_{2,i,t}$ are preset as follow: for $i\in\mathcal{N}_{\mathrm{S}}\cap\mathcal{N}_{\mathrm{G}}$, $a_{1,i}$=1 and $a_{2,i}$=1; for $i\in\mathcal{N}_{\mathrm{S}}\backslash\mathcal{N}_{\mathrm{G}}$, $a_{1,i}$=1 and $a_{2,i}$=0; for $i\in\mathcal{N}_{\mathrm{G}}\backslash\mathcal{N}_{\mathrm{S}}$, $a_{1,i}$=0 and $a_{2,i}$=1; and for $i\in\mathcal{N}\backslash(\mathcal{N}_{\mathrm{S}}\cup\mathcal{N}_{\mathrm{G}})$, $a_{1,i}$=0 and $a_{2,i}$=0. Constraints (9c)-(9d) ensure the power flow balance. Constraint (9e) indicates that a load that was already picked up cannot be de-energized. Constraint (9f) denotes the relationship between the voltage magnitudes of two adjacent nodes [8], [26]. Constraint (9g) bounds the voltage magnitude. We regard $V_{i,t}^2$ as a single variable; thus, (9f)-(9g) are linear. Constraint (9h) limits the power flow on each branch. Components without normal switches are considered by (9i), which restricts the connected/picked-up status of faulted branches/nodes [26]. Specifically, a load with a faulted closed load switch is picked up as long as the node where the load is located is energized. Repair resource dispatch is meaningful; nonetheless, it is beyond the scope of this paper, and relevant research can be found in [8]. Here, we assume that the repair process has already been known or anticipated in advance of scheduling and can be reflected as the shrinking sets $\mathcal{L}_{\mathrm{FO},t}$, $\mathcal{L}_{\mathrm{FC},t}$, $\mathcal{N}_{\mathrm{FO},t}$, $\mathcal{N}_{\mathrm{FC},t}$. For example, if a damaged branch $(i, j)$ is to be repaired during time span $t$, then $(i, j)$ is removed from $\mathcal{L}_{\mathrm{FO}}$ afterward. Constraint (9j) assumes that the substation node is always energized. In addition, if the substation loses power from the bulk grid, then (9j) is removed. Constraints (9k)-(9m) denote that any other node is energized if a Mod or an MEG arrives or it is connected to at least one energized node around it.

TABLE I
THE NUMBERS OF VARIABLES AND CONSTRAINTS OF THE PROPOSED MODEL

| Object of estimation | Number |
|---|---|
| Binary variables | $\lvert\mathcal{T}\rvert\cdot[\,3\cdot(\lvert\mathcal{M}_{\mathrm{G}}\rvert+\lvert\mathcal{M}_{\mathrm{F}}\rvert)\cdot(\lvert\mathcal{N}_{\mathrm{G}}\rvert+\lvert\mathcal{N}_{\mathrm{DP}}\rvert)+\lvert\mathcal{M}_{\mathrm{S}}\rvert\cdot\lvert\mathcal{N}_{\mathrm{S}}\rvert\cdot(\lvert\mathcal{K}\rvert+2)$ $+3\cdot\lvert\mathcal{N}_{\mathrm{S}}\rvert\cdot\lvert\mathcal{K}\rvert+\lvert\mathcal{M}_{\mathrm{S}}\rvert\cdot(\lvert\mathcal{K}\rvert+1)+(L+2)\cdot\lvert\mathcal{M}_{\mathrm{G}}\rvert+\lvert\mathcal{M}_{\mathrm{F}}\rvert+6\cdot\lvert\mathcal{L}\rvert$ $+4\cdot\lvert\mathcal{N}\rvert\,]+2\cdot\lvert\mathcal{M}_{\mathrm{S}}\rvert\cdot\lvert\mathcal{N}_{\mathrm{S}}\rvert+2\cdot(\lvert\mathcal{M}_{\mathrm{G}}\rvert+\lvert\mathcal{M}_{\mathrm{F}}\rvert)\cdot(\lvert\mathcal{N}_{\mathrm{G}}\rvert+\lvert\mathcal{N}_{\mathrm{DP}}\rvert)$ $+\lvert\mathcal{M}_{\mathrm{S}}\rvert+\lvert\mathcal{M}_{\mathrm{G}}\rvert+\lvert\mathcal{M}_{\mathrm{F}}\rvert+(\lvert\mathcal{M}_{\mathrm{S}}\rvert+\lvert\mathcal{N}_{\mathrm{S}}\rvert)\cdot\lvert\mathcal{K}\rvert$ |
| Continuous variables | $\lvert\mathcal{T}\rvert\cdot[\,(3\cdot\lvert\mathcal{M}_{\mathrm{G}}\rvert+\lvert\mathcal{M}_{\mathrm{F}}\rvert)\cdot(\lvert\mathcal{N}_{\mathrm{G}}\rvert+\lvert\mathcal{N}_{\mathrm{DP}}\rvert)+2\cdot\lvert\mathcal{M}_{\mathrm{G}}\rvert\cdot\lvert\mathcal{N}_{\mathrm{G}}\rvert+3\cdot\lvert\mathcal{N}_{\mathrm{S}}\rvert\cdot\lvert\mathcal{K}\rvert$ $+2\cdot\lvert\mathcal{M}_{\mathrm{S}}\rvert+3\cdot\lvert\mathcal{M}_{\mathrm{G}}\rvert+3\cdot\lvert\mathcal{M}_{\mathrm{F}}\rvert+\lvert\mathcal{N}_{\mathrm{G}}\rvert+\lvert\mathcal{N}_{\mathrm{DP}}\rvert+\lvert\mathcal{K}\rvert+(2\cdot\lvert\mathcal{L}\rvert+3)\cdot\lvert\mathcal{N}\rvert\,]$ $+2\cdot\lvert\mathcal{M}_{\mathrm{S}}\rvert+3\cdot\lvert\mathcal{M}_{\mathrm{G}}\rvert+3\cdot\lvert\mathcal{M}_{\mathrm{F}}\rvert+\lvert\mathcal{N}_{\mathrm{G}}\rvert+\lvert\mathcal{N}_{\mathrm{DP}}\rvert+\lvert\mathcal{K}\rvert$ |
| Constraints | $\lvert\mathcal{T}\rvert\cdot[\,(10\cdot\lvert\mathcal{M}_{\mathrm{G}}\rvert+8\cdot\lvert\mathcal{M}_{\mathrm{F}}\rvert)\cdot(\lvert\mathcal{N}_{\mathrm{G}}\rvert+\lvert\mathcal{N}_{\mathrm{DP}}\rvert)+5\cdot\lvert\mathcal{M}_{\mathrm{S}}\rvert\cdot\lvert\mathcal{N}_{\mathrm{S}}\rvert\cdot(\lvert\mathcal{K}\rvert+1)$ $+3\cdot\lvert\mathcal{M}_{\mathrm{S}}\rvert\cdot\lvert\mathcal{K}\rvert+9\cdot\lvert\mathcal{N}_{\mathrm{S}}\rvert\cdot\lvert\mathcal{K}\rvert+6\cdot\lvert\mathcal{M}_{\mathrm{G}}\rvert\cdot\lvert\mathcal{N}_{\mathrm{G}}\rvert+\lvert\mathcal{M}_{\mathrm{G}}\rvert\cdot\lvert\mathcal{N}_{\mathrm{DP}}\rvert+7\cdot\lvert\mathcal{M}_{\mathrm{S}}\rvert$ $+(4\cdot L+23)\cdot\lvert\mathcal{M}_{\mathrm{G}}\rvert+9\cdot\lvert\mathcal{M}_{\mathrm{F}}\rvert+12\cdot\lvert\mathcal{K}\rvert+3\cdot\lvert\mathcal{N}_{\mathrm{G}}\rvert+3\cdot\lvert\mathcal{N}_{\mathrm{DP}}\rvert+\lvert\mathcal{N}\rvert^2$ $+4\cdot\lvert\mathcal{L}\rvert\cdot\lvert\mathcal{N}\rvert+13\cdot\lvert\mathcal{N}\rvert+14\cdot\lvert\mathcal{L}\rvert+\Sigma_{t\in\mathcal{T}}(\lvert\mathcal{L}_{\mathrm{FO},t}\rvert+\lvert\mathcal{L}_{\mathrm{FC},t}\rvert+\lvert\mathcal{N}_{\mathrm{FO},t}\rvert$ $+\lvert\mathcal{N}_{\mathrm{FC},t}\rvert)-5\,]+7\cdot\lvert\mathcal{M}_{\mathrm{S}}\rvert+8\cdot\lvert\mathcal{M}_{\mathrm{G}}\rvert+8\cdot\lvert\mathcal{M}_{\mathrm{F}}\rvert+3\cdot\lvert\mathcal{K}\rvert+\lvert\mathcal{N}_{\mathrm{G}}\rvert+\lvert\mathcal{N}_{\mathrm{DP}}\rvert$ |

### *C. Linearization of Nonlinear Constraints*

Note that (5c), (6b), (9h), and (9l) are nonlinear. The first three can be simply linearized based on the method in [28] (see (13) in [28]). For the product term in (9l), *e.g.*, $\eta_{i',t}\cdot\kappa_{ii',t}$, according to [22], we introduce a new binary variable $\chi_{i'-ii',t}$ to replace $\eta_{i',t}\cdot\kappa_{ii',t}$ by adding the following constraints:

$$\chi_{i'-ii',t}\geq\eta_{i',t}+\kappa_{ii',t}-1\ ,\ \chi_{i'-ii',t}\leq\eta_{i',t}\ ,\ \chi_{i'-ii',t}\leq\kappa_{ii',t} \tag{10}$$

Thus, (9l) is equivalently linearized. The above method is actually identical to the McCormick envelopes [29] used in [26].

## IV. SCHEDULING MODEL TO BOOST RESILIENCE

We set the objective function as follows, which was commonly used as a good metric to assess the DS resilience under restoration [5], [8], [26], [30].

$$\begin{aligned}\max\sum\nolimits_{t\in\mathcal{T}}\Big[&\sum\nolimits_{i\in\mathcal{N}}w_i\delta_{i,t}P_{i,t}^{\mathrm{L}}\Delta t-\varphi_{travel}\Big(\sum\nolimits_{j\in\mathcal{M}_{\mathrm{S}}}\sum\nolimits_{i\in\mathcal{N}_{\mathrm{S}}}v_{j,i,t}\\&+\sum\nolimits_{m\in\mathcal{M}_{\mathrm{G}}}\sum\nolimits_{i\in\mathcal{N}_{\mathrm{G}}\cup\mathcal{N}_{\mathrm{dp}}}v_{m,i,t}+\sum\nolimits_{h\in\mathcal{M}_{\mathrm{F}}}\sum\nolimits_{i\in\mathcal{N}_{\mathrm{G}}\cup\mathcal{N}_{\mathrm{dp}}}v_{h,i,t}\Big)\\&-\varphi_{fuel}\Big(\sum\nolimits_{m\in\mathcal{M}_{\mathrm{G}}}\sum\nolimits_{i\in\mathcal{N}_{\mathrm{S}}}\iota_{m,i,t}+\sum\nolimits_{h\in\mathcal{M}_{\mathrm{F}}}\sum\nolimits_{i\in\mathcal{N}_{\mathrm{S}}}\iota_{h,i,t}\Big)\Big]\end{aligned} \tag{11}$$

The first term of (11) is the weighted sum of the picked-up energy of loads during scheduling. Two coefficients $\varphi_{travel}$ and $\varphi_{fuel}$ are introduced as penalties with regard to the traveling of MERs and the fuel exchange between MEGs/FTs and nodes, as in [8]. Specifically, the second term of (11), weighted by $\varphi_{travel}$, is intended to prevent futile travel of the MERs, and the third term, weighted by $\varphi_{fuel}$, is used to prevent redundant fuel exchange between FTs or MEGs and nodes. In general, as stated in [8], the values of $\varphi_{travel}$ and $\varphi_{fuel}$ can be set based on the decision-makers' preference or the analytic hierarchic process [31]. We can also adjust the values of $w_i$, $\varphi_{travel}$ and $\varphi_{fuel}$ based on an assessment of interruption cost of loads, travel cost of MERs, and labor cost for handling fuel exchange.

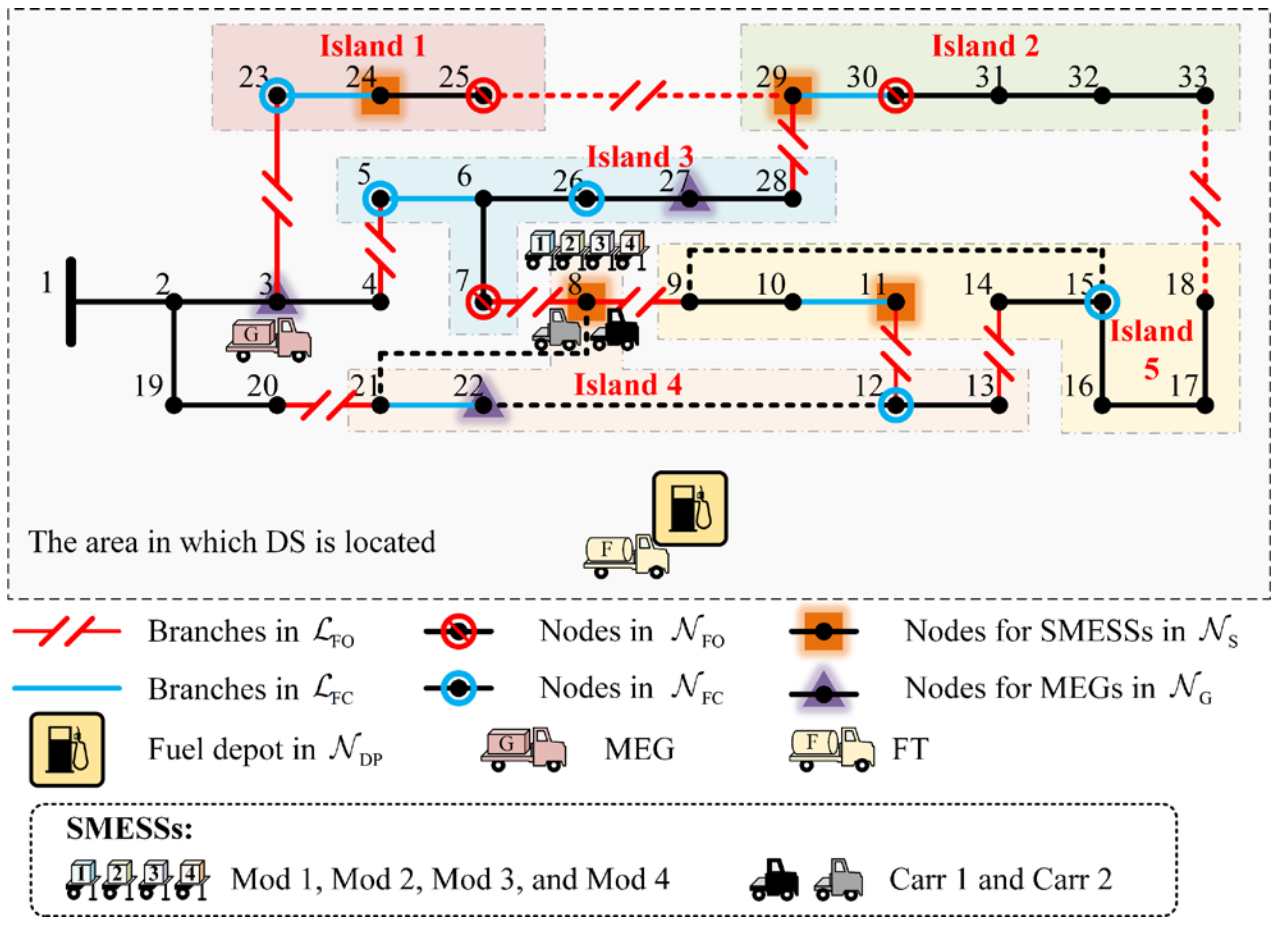

Fig. 3. IEEE 33-node test system before scheduling.

Thus, the scheduling model involving SMESSs, MEGs, FTs, and reconfiguration to boost the DS resilience is expressed as:

*Objective*: (11); *Constraints*: (1)-(9), along with (10).

The numbers of variables (both the binary and continuous ones) and constraints of the proposed model are analyzed, respectively. The estimation results are given in Table I. Specifically, in the estimation of constraints, a constraint in bilateral form (for example, $SOC_{k,\min} \le SOC_{k,t} \le SOC_{k,\max}$) is counted as two constraints, and the linearization methods described in Section III.C is considered.

Particularly, it is simple to prove that the traditional MESS solutions always lie in the feasible region of the above proposed model, and thus, the optimal SMESS solution has at least no worse objective value and performance than those of the MESS. Essentially, a traditional MESS is a special case of an SMESS by letting its Carr and some fixed Mods be always together. Suppose that an MESS comprises Carr 1 and Mods 1 & 2, and then, any scheduling solution of it can be represented as "when Carr 1 is traveling, $\gamma_{1,1,t}=1$ and $\gamma_{2,1,t}=1$; and when Carr 1 is parked at node $i$, $\zeta_{1,i,t}=1$ and $\zeta_{2,i,t}=1$" in our model. In other words, for an MESS solution, we can always find an equivalent SMESS solution that does not cause any violation of constraints (2)-(4) as well as the others. Thus, the MESS solutions are always included in the feasible region of the proposed model.

In addition, it takes some time for a Carr to drop or pick up some Mods at the location of a node. However, this time is so small that we can just ignore it in the scheduling. Or, we can also consider it in the model by easily adding this time to the parameter $T_{j,ii'}$. More generally, the parameter $T_{j,ii'}$ can truly represent the whole time spent in an actual travel of Carr $j$ (and for other MERs) from nodes $i$ to $i'$, including the time for traveling, for dropping and picking up Mods, and even for electrically connecting Mods to and disconnecting them from an DS node and for some others. We just add them up and treat the sum as the new $T_{j,ii'}$ to consider them in the scheduling.

TABLE II
THE SETS OF FAULTS $\mathcal{N}_{FO,t}$, $\mathcal{N}_{FC,t}$, $\mathcal{L}_{FO,t}$, AND $\mathcal{L}_{FC,t}$ DURING SCHEDULING

| Faults | $\mathcal{N}_{FO,t}$ | $\mathcal{N}_{FC,t}$ | $\mathcal{L}_{FO,t}$ | $\mathcal{L}_{FC,t}$ |
|---|---|---|---|---|
| **1st-6th** time spans | Nodes 7, 25, 30 | Nodes 5, 26, 12, 23, 15 | Branches (8,9), (11,12), (13,14), (20,21), (4,5), (7,8), (25,29), (3,23), (18,33), (28,29) | Branches (5,6), (23,24), (21,22), (29,30), (10,11) |
| **7th-11th** time spans | Φ | Nodes 12, 23, 15 | Branches (4,5), (7,8), (25,29), (3,23), (18,33), (28,29) | Branches (29,30), (10,11) |
| **12th** time span | Φ | Φ | Φ | Φ |

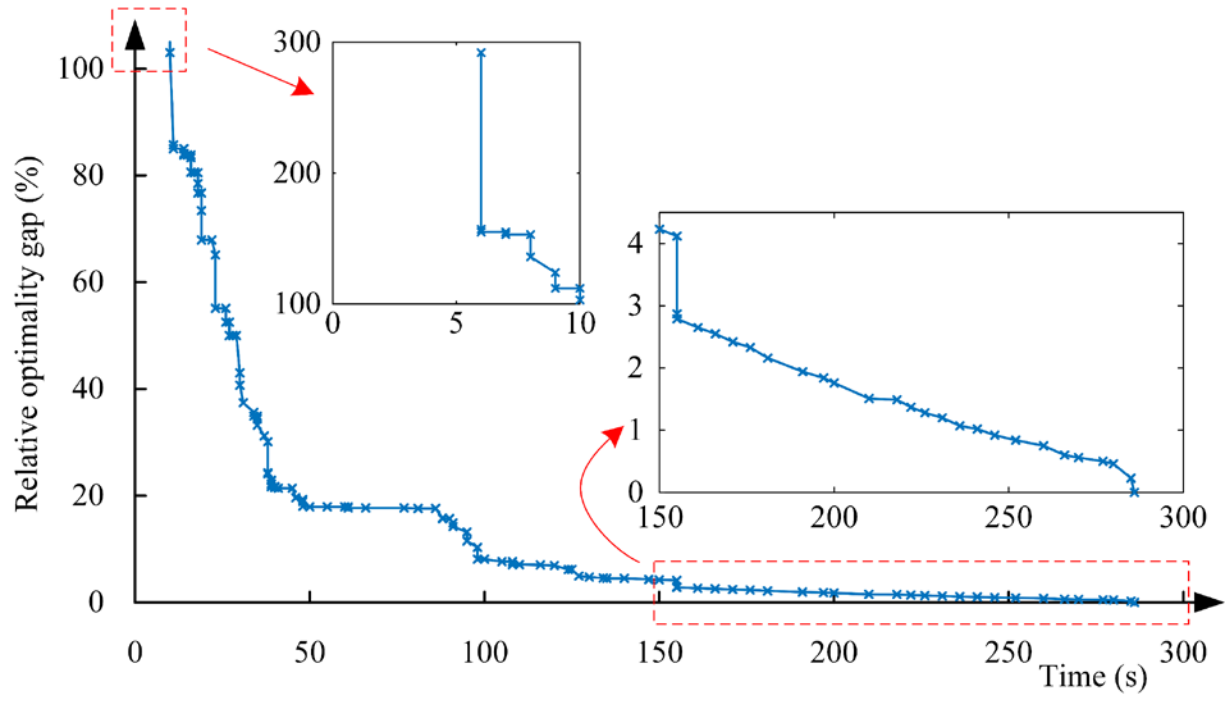

Fig. 4. Converging process of the relative optimality gap in 33-node test.

## V. NUMERICAL RESULTS

In this section, we conduct test and case studies to verify the effectiveness of the proposed model. The model is coded on the MATLAB R2020b platform with the YALMIP toolbox [32] and solved by Gurobi v9.1.2 on a laptop with an Intel Core i7-11800H 2.3 GHz CPU processor and 16 GB RAM.

### *A. Test A: IEEE 33-node System*

#### *1) Test System*

The modified IEEE 33-node system for test is shown in Fig. 3. Specific data, including impedance of branches and loads at nodes, can be found in [27]. Due to the lack of relevant data, we simply assume that the priority weights of loads are randomly assigned from 1 to 5. The first term of the objective function in (11) is calculated in kW·h. In practice, reaching a high level of electric service restoration rapidly is clearly a primary task and much more urgent than the others for the DS operator in postdisaster conditions; hence, we assign a small value of 0.1 to $\varphi_{travel}$ and $\varphi_{fuel}$. We do this also to focus fully on and display the performance of the SMESS in enhancing DS resilience by alleviating the restriction of the other costs in the objective function. The base power and voltage in the test are set as 1 MW and 12.66 kV, respectively.

Four Mods, each with a 500 kW/1 MW·h capacity [16], and two tractors as Carrs are scheduled as SMESSs in the test. Each Carr can carry at most two Mods. The charging/discharging efficiency of each Mod is set as 0.95/0.95, and the allowable SOC range is 0.1-0.9. In addition, an 800 kW/1000 kVA MEG is used, with a fuel consumption rate of 64.4 L, 109.8 L, 155.2

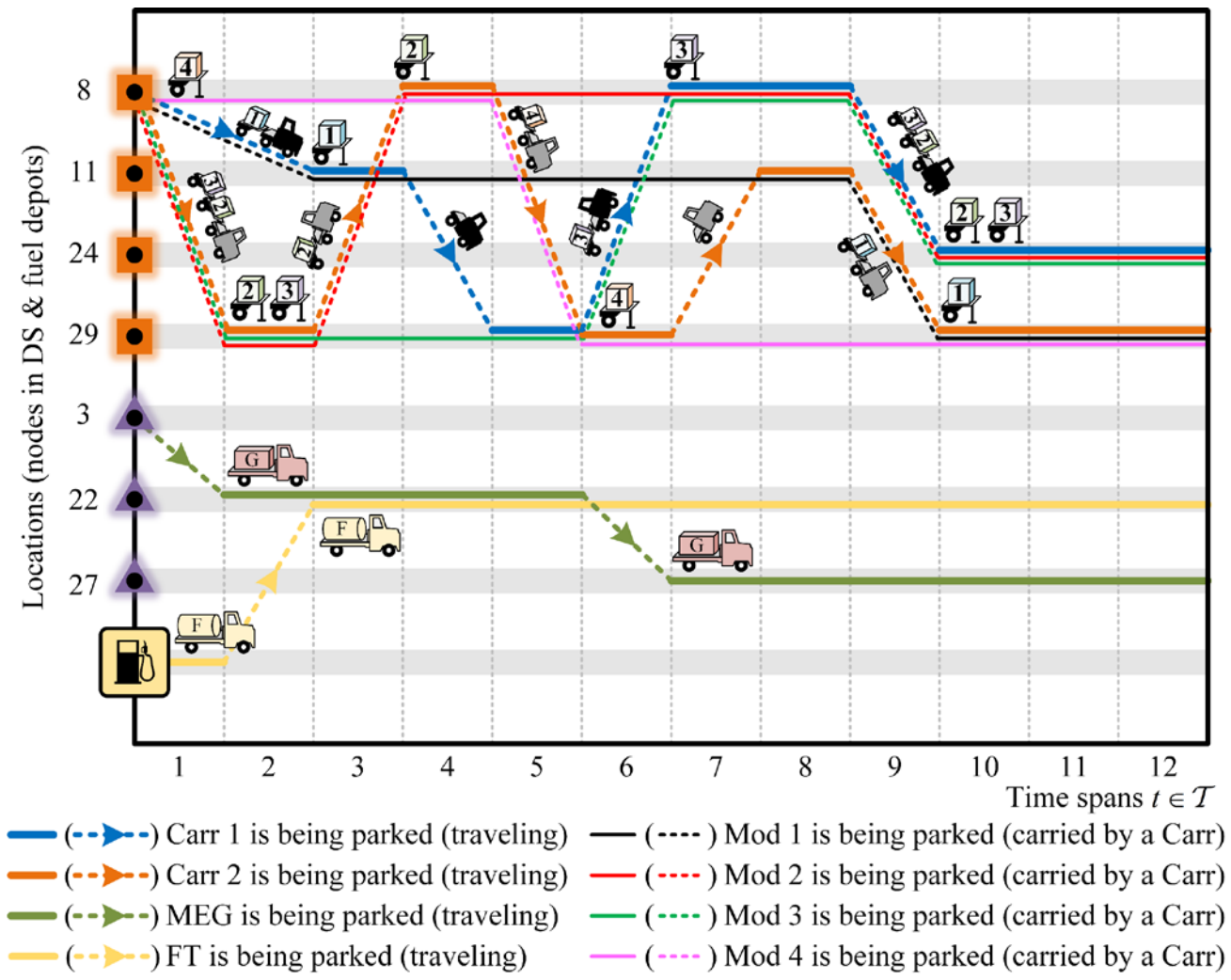

Fig. 5. Results of time-spatial behaviors of SMESSs, MEG, and FT.

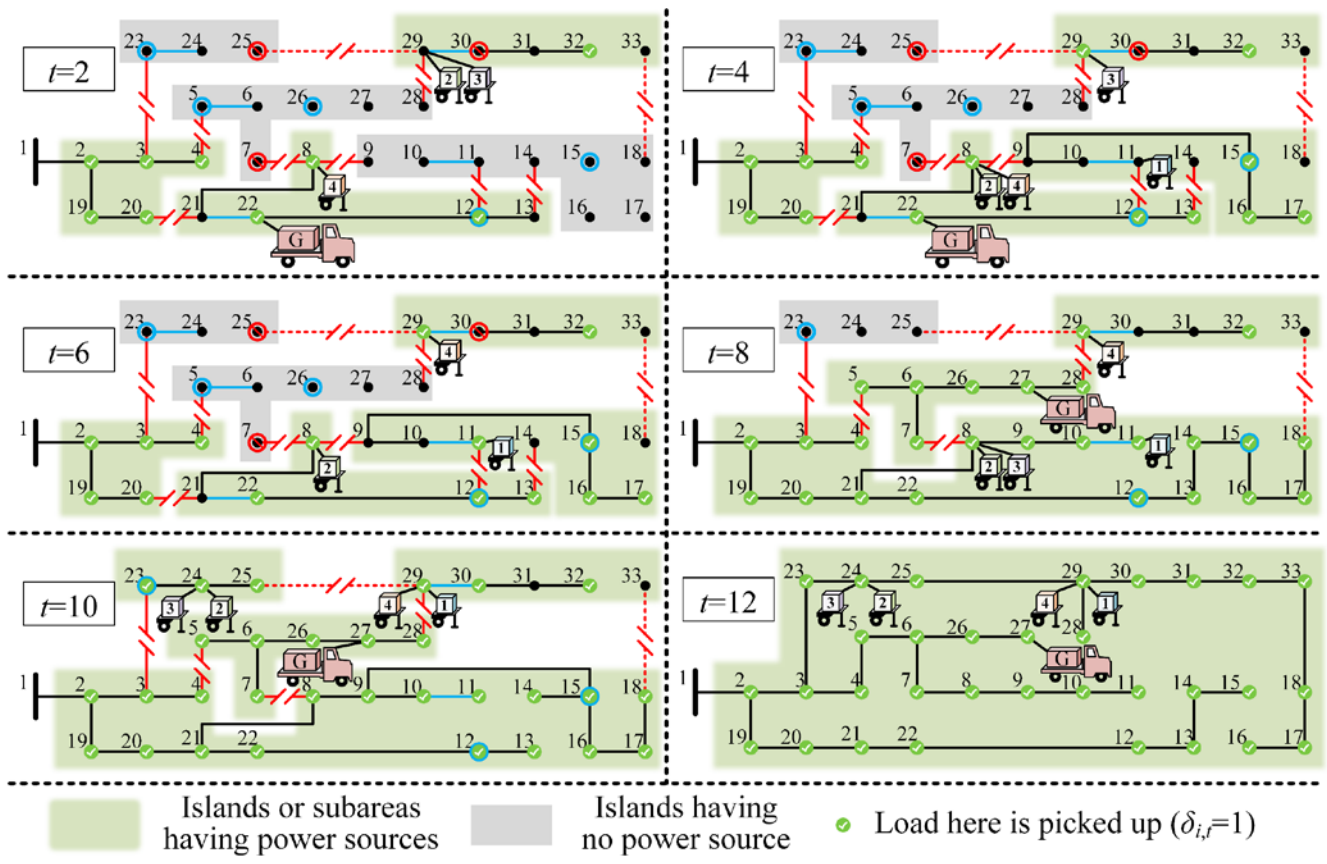

Fig. 6. Main results of DS reconfiguration and the picked-up states of loads.

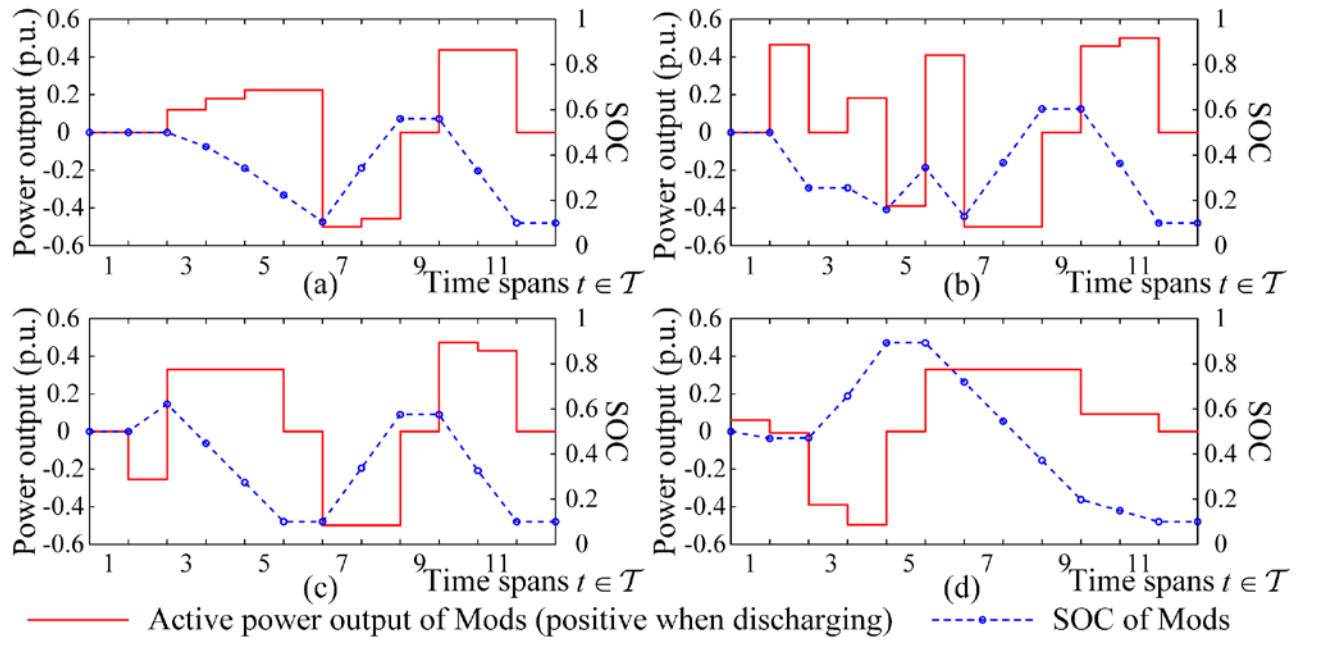

Fig. 7. Results of the active power output and SOC of (a) Mod 1, (b) Mod 2, (c) Mod 3, and (d) Mod 4.

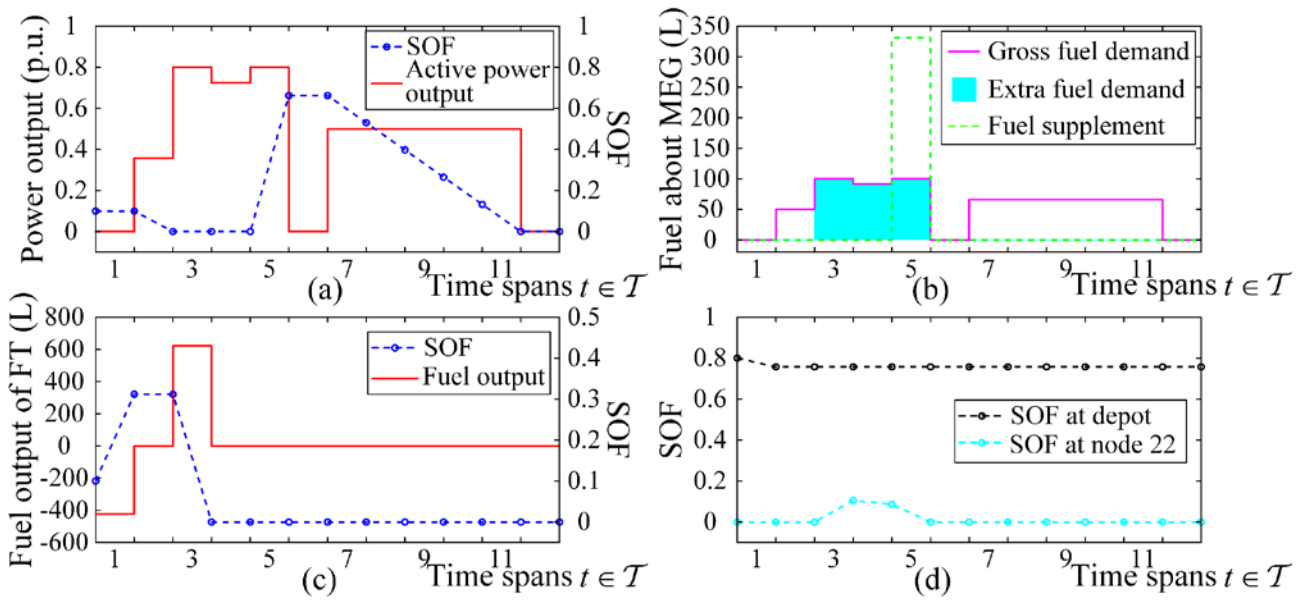

Fig. 8. Results of (a) active power output, SOF, (b) fuel demand/supplement of MEG; (c) fuel output and SOF of FT; and (d) SOF of node 22 and the depot.

L, and 200.6 L per hour at 1/4, 1/2, 3/4, and full load, respectively [24]. We assume that the MEG itself is equipped with a small 500 L fuel tank. A 2000 L FT also participates in the scheduling. The available fuel storage capacities of each node and the depot are set as 5000 L and 10000 L, respectively.

As shown in Fig. 3, four nodes {8, 11, 24, 29} support the connection of SMESSs to the DS, while three nodes {3, 22, 27} and the depot support the connection of the MEG/FT. For simplicity, we assume that all the Carrs, MEG and FT have the same speed, *i.e.*, spend the same time traveling between two locations, and that the time spent traveling from one to another and back are the same. Further, we arbitrarily assume the distances among nodes as follows: for {8, 11, 24, 29}, the time spans spent traveling between nodes 8 and 11, between 11 and 24, and between 24 and 29 are set as 2, while that between the other pairs is 1; for {3, 22, 27, depot}, the time spans spent traveling between nodes 3 and 27, between 3 and the depot, and between 27 and the depot are set as 2, while the others are 1.

*2) Scenario*

In the test, scheduling is performed over 6 h with a time step of $\Delta t$=0.5 h, *i.e.*, there is a total of 12 time spans. We consider a postdisaster scenario with a total of 23 faults, as depicted in Fig. 3, by which the DS is separated into 5 de-energized islands. The repair of faults is simply considered in a two-step way. As indicated by Table II, some of the faults are cleared after the 6th time span, and the others are cleared after the 11th time span.

All the Mods and Carrs of the SMESSs are initially located at node 8, and the initial SOC of each Mod is set as 0.5. The MEG and FT are initially located at node 3 and the depot, respectively, and both have the same initial SOF of 0.1. We set 0.8 as the initial SOF of the depot and 0 for all the nodes of the DS to represent the extreme condition where the DS operator has not prepared fuel within the DS for the MEG operation before disasters, which might often occur due to the poor predictability of many disasters.

*3) Simulation Results*

The tolerance *MIPGap* of the solver is set as 0.1%. The model is solved in 298 s, 285.6 s of which is used by the solver Gurobi and the rest is used by YALMIP and MATLAB to deal with the code. The relative optimality gap gradually converges to the final value 0, as shown in Fig. 4. By solving the proposed model, the main results are given by Fig. 5 to Fig. 8, which can verify the effectiveness of the proposed model. For clarity, we describe the main scheduling process from the perspectives of the five islands in Fig. 3.

We look at *Island 4* first, which includes the initial positions of SMESSs. As soon as scheduling begins, Mod 1 is transported by Carr 1 to node 11 (*Island 5*), Mods 2 and 3 are transported by Carr 2 to node 29 (*Island 2*), and Mod 4 stays at node 8 for a while. In addition, the MEG immediately moves from node 3 to node 22. In the 1st time span (*i.e.*, in $t$=1), Mod 4 discharges slightly to energize a small part of the loads in *Island 4*, and as the MEG arrives, which is of high power and energy capacity (the FT comes soon to supplement the fuel), Mod 4 begins to charge until it is transported by Carr 2 to node 29 in $t$=5. The MEG acts as the main source of *Island 4* until it leaves in $t$=6; before that, *i.e.* in $t$=5, Mod 2 charges intensively to store enough energy to then sustain the power supply of *Island 4* after the MEG leaves away. In $t$=7, the connection of Island 4 to the substation is recovered by the repair of faulted lines. Then, Mods 2 and 3 charge at the maximum power for two time spans and are sent by Carr 1 to node 24 in $t$=9.

For *Island 1*, as shown in Fig. 5, no power source is available until Mods 2 and 3 are transported by Carr 1 to node 24 in $t$=10.

For *Island 3*, it obtains a power supply when the MEG arrives at node 27 from node 22 in $t$=7, and all the loads in it are energized.

For *Island 5*, when scheduling begins, Carr 1 transports Mod 1 from node 8 to node 11. After it arrives, Mod 1 keeps discharging to supply part of the loads of *Island 5* until $t$=7 when *Island 5* is reconnected to the substation; then, Mod 1 begins to charge intensively for two time spans. In $t$=9, Mod 1 is carried by Carr 2 and transported to node 29.

For *Island 2*, it obtains a power supply soon after scheduling begins due to the timely arrival of Mods 2 and 3 carried by Carr 2. It is interesting to note that after arriving, Mod 3 charges in $t$=2 rather than discharging to provide some power supply. This is similar to that described above, where Mod 2 in *Island 4* charges intensively before the MEG leaves. Since Mod 2 is going to be carried away from *Island 2* in $t$=3, before that, Mod 3 "absorbs" enough energy from Mod 2 to sustain the power supply to *Island 2* after Mod 2 leaves. Then, in $t$=6, Mod 4 is brought by Carr 2, and Mod 3 is transported to node 8 by Carr 1. Subsequently, the power supply is held by Mod 4 during the following several time spans, until Mod 1 is brought by Carr 2 and joins in.

Regarding the fuel delivery, as shown in Fig. 5 and Fig. 8, the FT refuels itself adequately at the depot when scheduling begins. After that, the FT moves to node 22 and releases all the fuel there in $t$=3, resulting in an increase in the SOF of node 22. Thus, the MEG obtains extra fuel for power generation to supply *Island 4*, even though it has uses up the fuel stored in itself after $t$=2. Then, in $t$=5, the MEG fully refuels itself from the fuel at node 22 (see Fig. 8 (a), which shows a large increase in the SOF of the MEG), and then, in $t$=6, it moves to node 27 to energize *Island 3* for a relatively long duration.

*4) Case studies and Comparison*

Based on the above, the following five cases are studied to verify the advantage of the proposed method in boosting DS resilience over other measures.

*Case 1*: No energy storage or generator is available.

*Case 2*: Only stationary energy storage systems and generator are available, and scheduling of the FT is considered. Specifically, we fix the four Mods and the MEG at node 8 and node 3, respectively, during the whole scheduling horizon.

*Case 3*: Two general MESSs, each with a 1 MW/2 MW·h capacity (to ensure an equivalent capacity of energy resources to that of *Case 5*), and the MEG is available. Scheduling of the FT is also considered.

*Case 4*: The SMESSs and MEG are available, but scheduling of the FT is forbidden. However, in this case, we allow the MEG to refuel itself by going to the depot.

*Case 5*: The SMESSs, MEG, and FT are jointly scheduled, as we highlight in this paper and did in the previous test.

The revisions to the proposed model required to realize

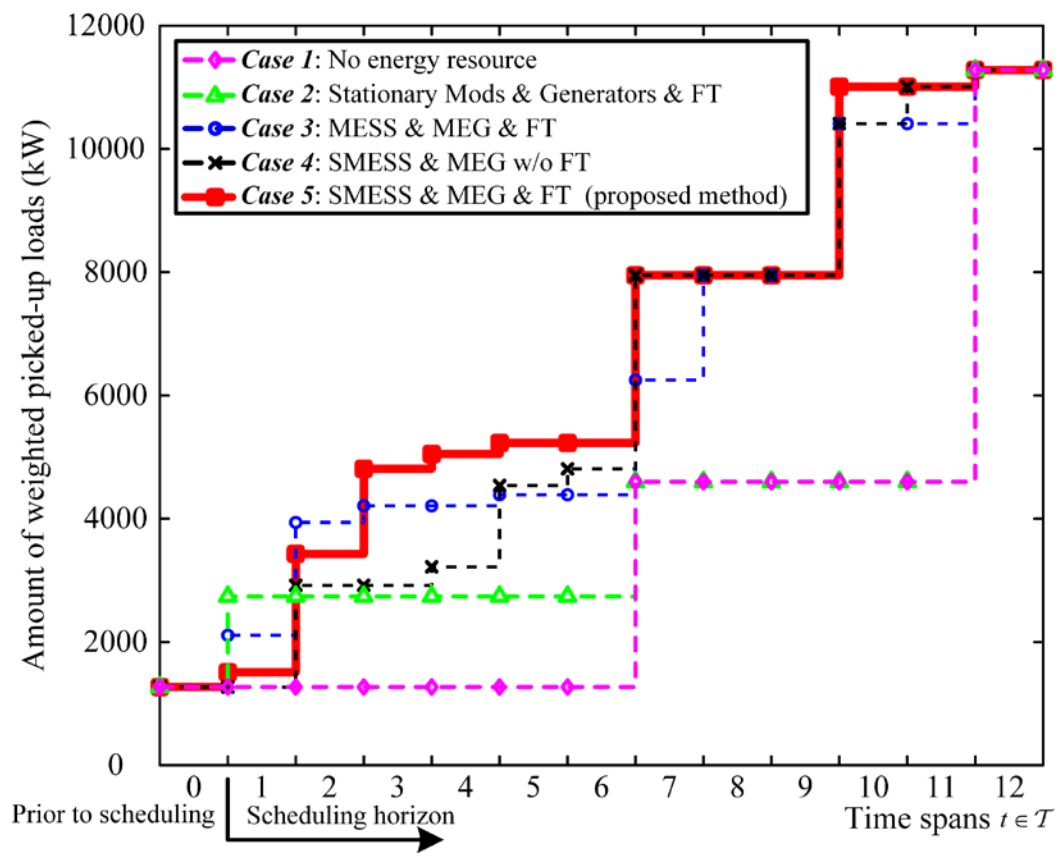

Fig. 9. Comparison among *Cases 1* to *5*.

*Cases 1* to *4* are given in the electronic appendix for this paper [33]. By solving the models, the comparison among the five cases regarding the performance in boosting DS resilience is shown in Fig. 9. Clearly, energy resources, whether stationary or with mobility, can give the DS better resilience. As shown in Fig. 9, the line of *Case 1* always lies below that of the others. Stationary energy resources can maintain the power supply to some loads after a blackout; however, they cannot serve the de-energized loads electrically isolated from them. As shown in Fig. 9, in the last several time spans, the performance of *Case 2* is reduced to that of *Case 1* since the stationary Mods and MEG cannot supply power to *Islands 1* to *3*, which are isolated from any power source by faulted lines. With mobility, energy resources can further boost the DS resilience to a higher level, as shown when comparing *Case 2* with *Case 3*/*4*/*5*, the latter of which have much higher lines in Fig. 9 than the former. In particular, *Case 5* results in a higher overall resilience level than that of *Case 3*. This is mainly because, in brief, as extensive faults cause multiple islands, the number of which could exceed the number of power sources (*e.g.*, there are 5 islands but 2 MESSs and 1 MEG in *Case 3*), the energy resources might become overwhelmed when attempting to serve all the de-energized islands. The proposed SMESS method could release and decouple the carrier and energy resources of the MESS (*e.g.*, we have more individual energy resources in *Case 5*, even though some of them have a smaller capacity than in *Case 3*) and thus allow the resources to be allocated in a more flexible and overall manner to serve more areas electrically isolated from each other. In addition, the scheduling of FTs in parallel with the scheduling of MEGs could offer better performance in terms of boosting the resilience, as shown when comparing *Case 4* and *Case 5*, since the fuel shortage MEGs might encounter could be mitigated in time and MEGs could go straight to the de-energized areas without too much consideration regarding whether to refuel themselves.

*5) Discussion about the Weight Selection*

As mentioned before, the weight coefficients $\varphi_{travel}$ and $\varphi_{fuel}$ are set to 0.1 in the previous test. To analyze the effect of the values of these two coefficients, an additional case study is conducted. We have assigned different values to $\varphi_{travel}$ and $\varphi_{fuel}$, from small to large, to analyze the resultant optimal solution after solving the corresponding model. In addition, for simplicity, the values of $\varphi_{travel}$ and $\varphi_{fuel}$ are kept equal during the above process. After a series of tests, the results are given in Table III. The previous test is referred as the base case.

TABLE III
RESULTS OF THE TERMS IN THE OBJECTIVE FUNCTION UNDER DIFFERENT VALUES OF $\varphi_{travel}$ AND $\varphi_{fuel}$

| $\varphi_{travel}$, $\varphi_{fuel}$ | **The 1st term**: weighted sum of the picked-up energy of loads (kW·h) | **The 2nd term**: total time spans for traveling | | | **The 3rd term**: total time spans for fuel exchange | |
|---|---|---|---|---|---|---|
| | | Carr | MEG | FT | MEG | FT |
| $1\times10^{-3}$ | 41025 | 10 | 2 | 1 | 1 | 2 |
| $1\times10^{-2}$ | 41025 | 10 | 2 | 1 | 1 | 2 |
| **$1\times10^{-1}$ (the base case)** | **41025** | **10** | **2** | **1** | **1** | **2** |
| 1 | 41025 | 10 | 2 | 1 | 1 | 2 |
| $1\times10^{1}$ | 41025 | 10 | 2 | 1 | 1 | 2 |
| $1\times10^{2}$ | 41025 | 10 | 2 | 1 | 1 | 2 |
| $1\times10^{3}$ | 38295 | 2 | 2 | 1 | 1 | 2 |
| $2\times10^{3}$ | 30385 | 2 | 0 | 0 | 0 | 0 |
| $5\times10^{3}$ | 25360 | 0 | 0 | 0 | 0 | 0 |
| $1\times10^{4}$ | 25360 | 0 | 0 | 0 | 0 | 0 |

From Table III, the general effect of weight selection in the objective function can be analyzed as follows. Obviously, the objective function in (11) shows a trade-off between restoring the de-energized loads and saving the cost of deploying the resources (*i.e.*, driving MERs and exchanging fuel), and the latter is weighted by $\varphi_{travel}$ and $\varphi_{fuel}$.

When the value of $\varphi_{travel}$ and $\varphi_{fuel}$ ranges from $1\times10^{-3}$ to $1\times10^{2}$, the optimal results regarding the behaviors of SMESSs, MEG, and FT are not changed, as shown in Table III, and the optimal behaviors shown in Fig. 5 is maintained. It means that, in this range of $\varphi_{travel}$ and $\varphi_{fuel}$, fully deploying the available resources (*i.e.*, those MERs) to restore the loads contributes more to the objective function than deploying them in a cost-saving way, and thus, the former is preferred. In other words, the resultant optimal decision aims to restore the loads by fully deploying the available MERs. Thus, in this range of $\varphi_{travel}$ and $\varphi_{fuel}$, the weighted sum of picked-up energy of loads (*i.e.*, the 1st term of the objective function) has the highest value compared with the other ranges, as shown in Table III. In addition, in this range, all of the MERs have been utilized to the greatest degree while futile deployment is prevented, and the decrease of weights $\varphi_{travel}$ and $\varphi_{fuel}$ (even to 0.001) will not cause any result of more frequent deployment of MERs.

Then, the value of $\varphi_{travel}$ and $\varphi_{fuel}$ is increased to $1\times10^{3}$, and things are different from the above. The increased weights make the cost of resource deployment contribute so much to the objective function that it becomes comparable to the restored energy of loads and can no longer be ignored. Thus, in the resultant optimal decision, the MERs are deployed in a more conservative way to take care of the cost of resource deployment. Compared with 10 time spans before, the Carrs of SMESSs are only scheduled to travel for 2 time spans now, as shown in Table III: Specifically, two Mods are carried by one Carr from node 8 to node 29 during time span 4 and the other two Mods are carried by another Carr from node 8 to node 24 during time span 9. In addition, even though the weaker utilization of SMESSs reduces the cost of resource deployment, it also leads to the decrease of the restored energy of loads.

Similarly, when $\varphi_{travel}$ and $\varphi_{fuel}$ are increased to $2\times10^{3}$, the MERs are deployed more conservatively and less frequently. The MEG and FT are even not deployed in this case, and this further decreases the restored energy of loads to a lower level.

Then, when $\varphi_{travel}$ and $\varphi_{fuel}$ are further increased to $5\times10^{3}$ or higher, it costs too much to deploy the MERs among the DS to restore the de-energized loads, and all of the MERs stay at their initial locations. The loads can only be picked up by the local power sources (*e.g.*, the Mods in the same island) or by the substation as long as they are reconnected to it due to repair of fault components. Accordingly, the restored energy of loads is decreased to the lowest level.

Actually, the case of a very high value of $\varphi_{travel}$ and $\varphi_{fuel}$ larger than 100 is commonly not realistic, if we note the interruption cost of loads and the cost of resource deployment. For example, according to [7], the unit interruption cost of non-critical loads was set to \$2/kW·h (which can be seen as the profit for restoring 1 kW·h energy of loads), and the unit cost for MERs traveling was set to \$80/h, *i.e.*, \$40 for traveling for one time span, which is 0.5 h in our case study. Based on these, the value of $\varphi_{travel}$ and $\varphi_{fuel}$ should be set to 20. Thus, except for special requirement (*e.g.*, a very high cost for deploying MERs is required), the value of $\varphi_{travel}$ and $\varphi_{fuel}$ selected by the decision-makers may well lie in the range of $1\times10^{-3}$ to $1\times10^{2}$. Thus, based on the above analysis, the results including Fig. 5 can be representative of the general circumstances where the weights $\varphi_{travel}$ and $\varphi_{fuel}$ are selected in the reasonable range.

## B. *Test B: IEEE 123-node System*

### 1) *Test System*

To verify the scalability of the proposed model, a further test is conducted on the larger IEEE 123-node system, as shown in Fig. 10. Specific data about the test system can be found in [34]. The priority weights of loads are randomly assigned from 1 to 5 and both of $\varphi_{travel}$ and $\varphi_{fuel}$ are set to 0.1, as we did in Test A. The base power and voltage are 1 MW and 4.16 kV, respectively. In this test, the SMESSs to be scheduled comprise six Mods and three Carrs. In addition, two MEGs and one FT are considered. All of them have the same parameters as in Test A in terms of capacity (regarding power, energy, fuel storage, and carrying load of Carr), charging and discharging efficiency, fuel consumption rate, etc.

In the test system of Fig. 10, suppose that four nodes {15, 54, 59, 104} can support the connection of SMESSs to the DS and three nodes {27, 70, 83}, along with the fuel depot, support the connection of MEGs and FT. The travel time between nodes is arbitrarily assumed as follows: for {15, 54, 59, 104}, a Carr will spend two time spans traveling between nodes 15 and 54 or between nodes 15 and 104, while one time span is required for moving between the other pairs of nodes; for {27, 70, 83, depot}, an MEG or FT will spend two time spans traveling between nodes 27 and 83 and one time span traveling between the other pairs.

### 2) *Scenario*

The scheduling is performed over 6 h with a time step $\Delta t$=0.5 h. Suppose that 57 faults have occurred within the DS after a major disaster, as expressed in Table IV, and they separate the

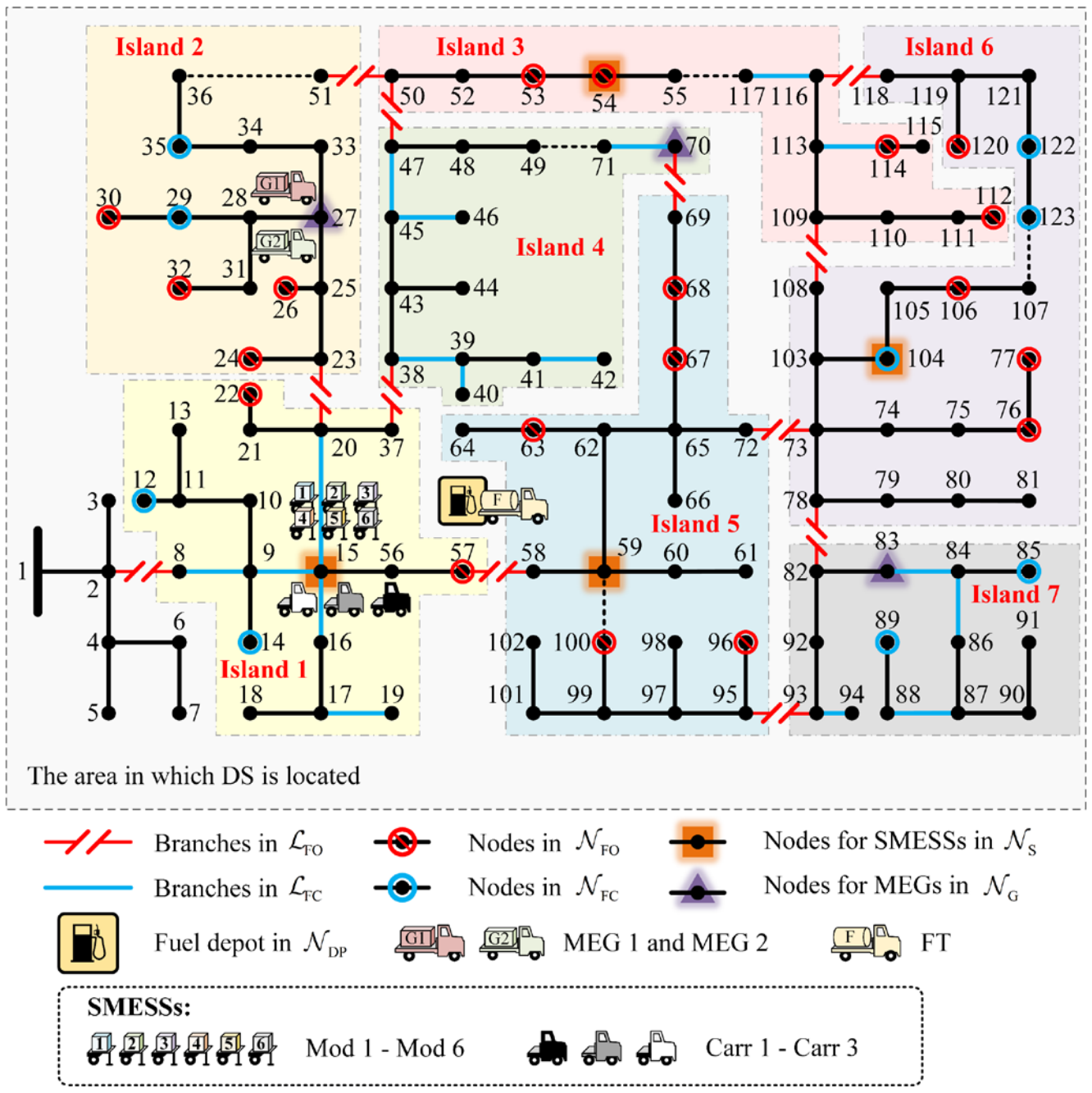

Fig. 10. IEEE 123-node test system before scheduling.

TABLE IV
THE SETS OF FAULTS $\mathcal{N}_{FO,t}$, $\mathcal{N}_{FC,t}$, $\mathcal{L}_{FO,t}$, AND $\mathcal{L}_{FC,t}$ DURING SCHEDULING

| Faults | $\mathcal{N}_{FO,t}$ | $\mathcal{N}_{FC,t}$ | $\mathcal{L}_{FO,t}$ | $\mathcal{L}_{FC,t}$ |
|---|---|---|---|---|
| **1st-6th** time spans | Nodes 22, 57, 24, 26, 30, 32, 53, 54, 112, 114, 63, 67, 68, 96, 100, 76, 77, 106, 120 | Nodes 12, 14, 29, 35, 85, 89, 104, 122, 123 | Branches (2,8), (20,23), (37,38), (50,51), (47,50), (69,70), (116,118), (108,109), (72,73), (78,82), (93,95), (57,58) | Branches (8,9), (15,16), (39,40), (45,47), (70,71), (84,86), (116,117), (15,20), (9,15), (17,19), (38,39), (41,42), (45,46), (83,84), (87,88), (93,94), (113,114) |
| **7th-11th** time spans | Nodes 114, 63, 67, 68, 96, 100, 76, 77, 106, 120 | Φ | Branches (69,70), (116,118), (108,109), (72,73), (78,82), (93,95), (57,58) | Branches (15,20), (9,15), (17,19), (38,39), (41,42), (45,46), (83,84), (87,88), (93,94), (113,114) |
| **12th** time span | Φ | Φ | Φ | Φ |

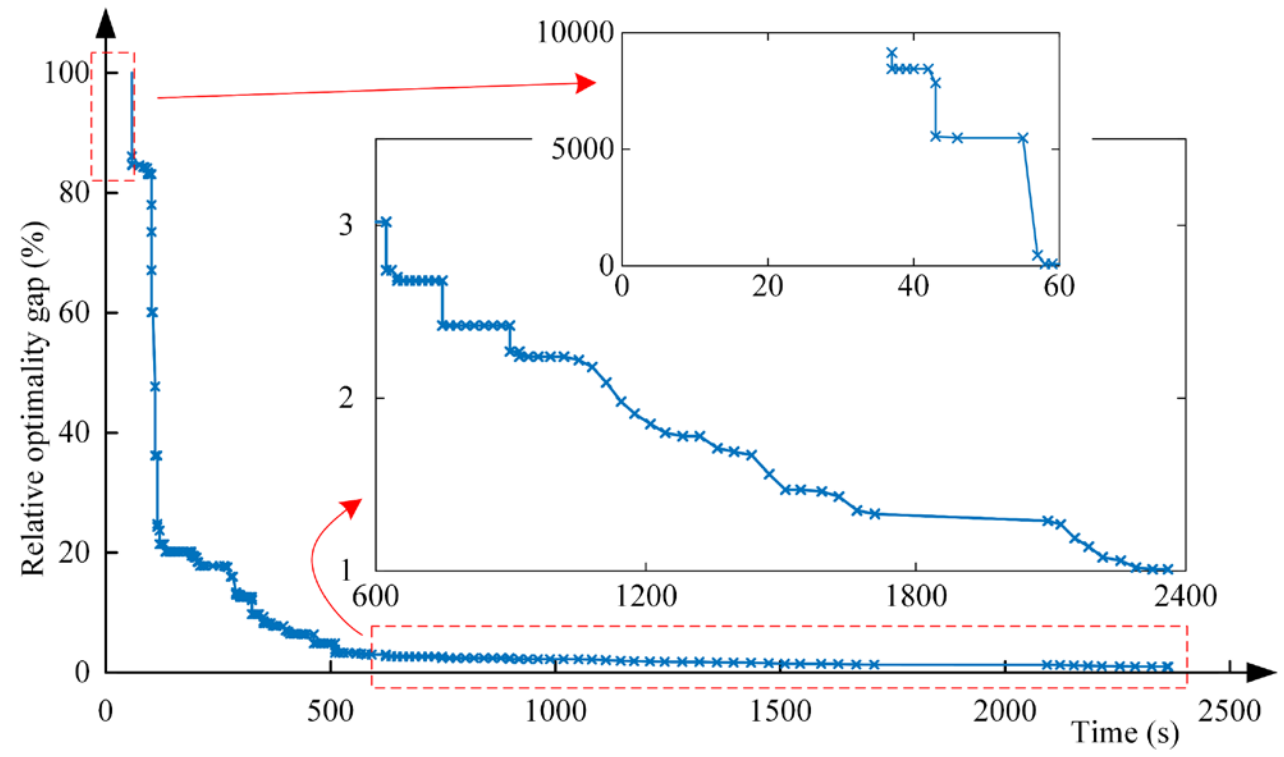

Fig. 11. Converging process of the relative optimality gap in 123-node test.

DS into 7 de-energized islands that lost continuous power supply from the substation. A two-step repair of the faults is also simply considered, as we did before. The SMESSs, MEGs and FT are initially located at node 15, node 27, and the depot, respectively. In addition, all the Mods have the same initial SOC of 0.5, and the MEGs/FTs, the depot and nodes {27, 70,

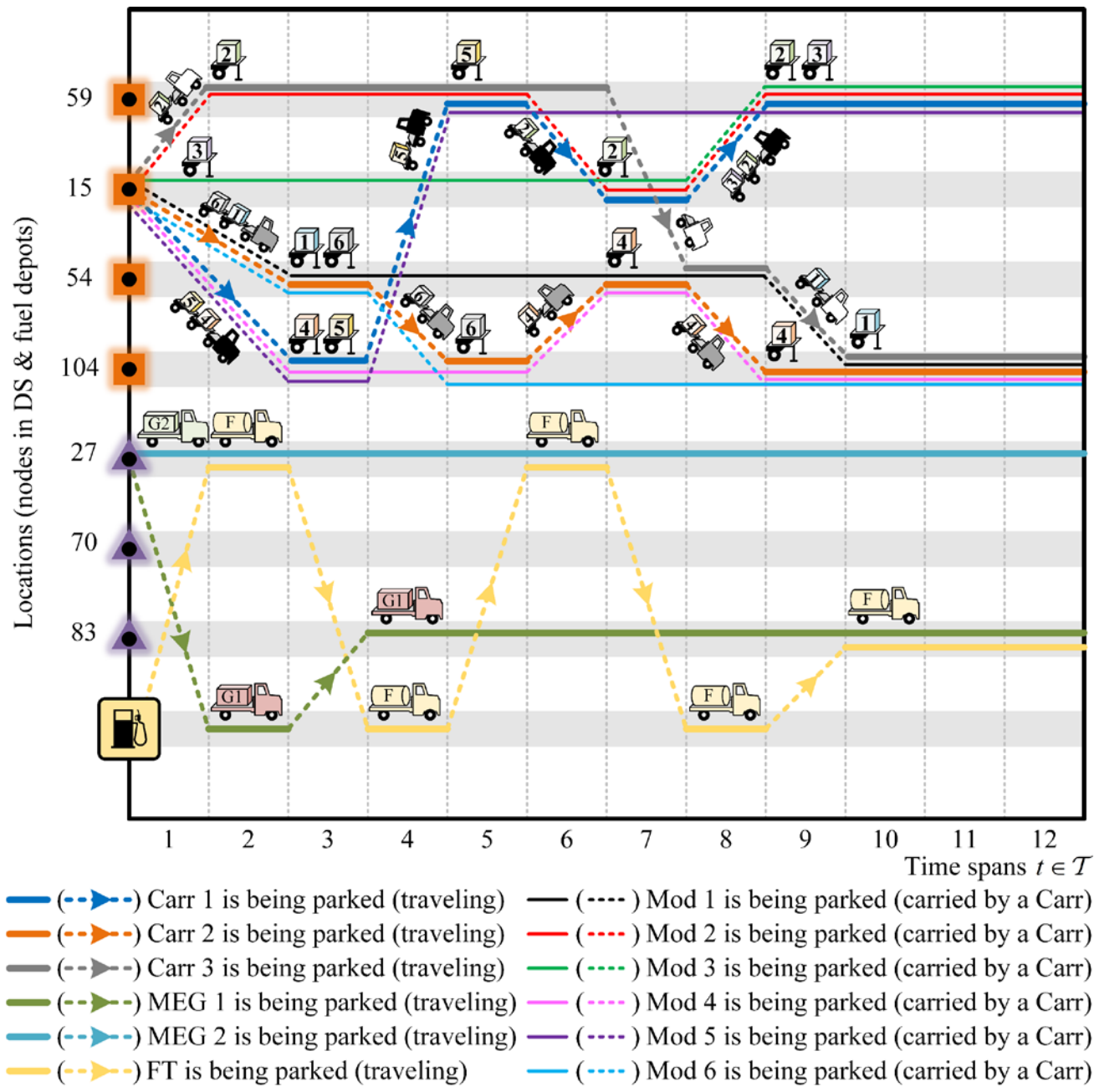

Fig. 12. Results of time-spatial behaviors of SMESSs, MEGs, and FT.

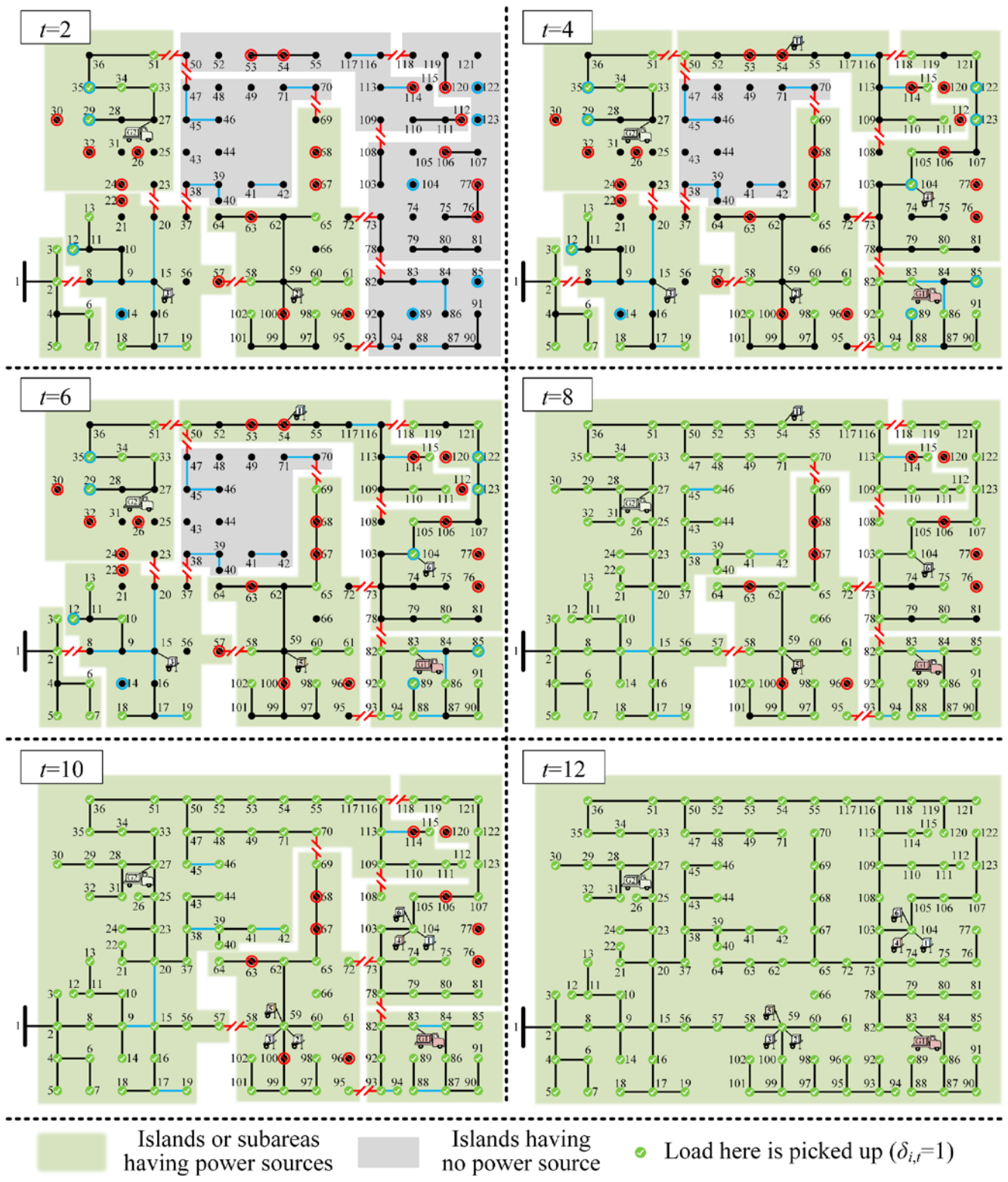

Fig. 13. Main results of DS reconfiguration and the picked-up states of loads.

83} have the initial SOF of 0.1, 0.8 and 0, respectively.

*3) Simulation Results*

The tolerance *MIPGap* is set as 1%. The model is solved in 2589 s. 2363 s of this time is spent by the solver Gurobi and the others by YALMIP and MATLAB. The converging process of the relative optimality gap is shown in Fig. 11. Some measures can be considered to save the time to obtain the solution. For example, from Fig. 11, we see that the optimality gap has dropped below 3% after about 10 minutes and a further drop takes so long. Thus, we can truncate the converging process and advance the termination of the solver by slightly enlarging the tolerance (*i.e.*, the optimality gap below which the solver should terminate, like *MIPGap* for Gurobi) to the degree that can be acceptable in practical engineering. By doing this, even though the solution may be not as good as the one under a smaller tolerance gap, it may significantly reduce the computation time at a slight sacrifice of the solution quality. In addition, even though solving a model on different computers suffers performance variability, a stronger computer equipped with faster CPU and memory commonly also helps [35].

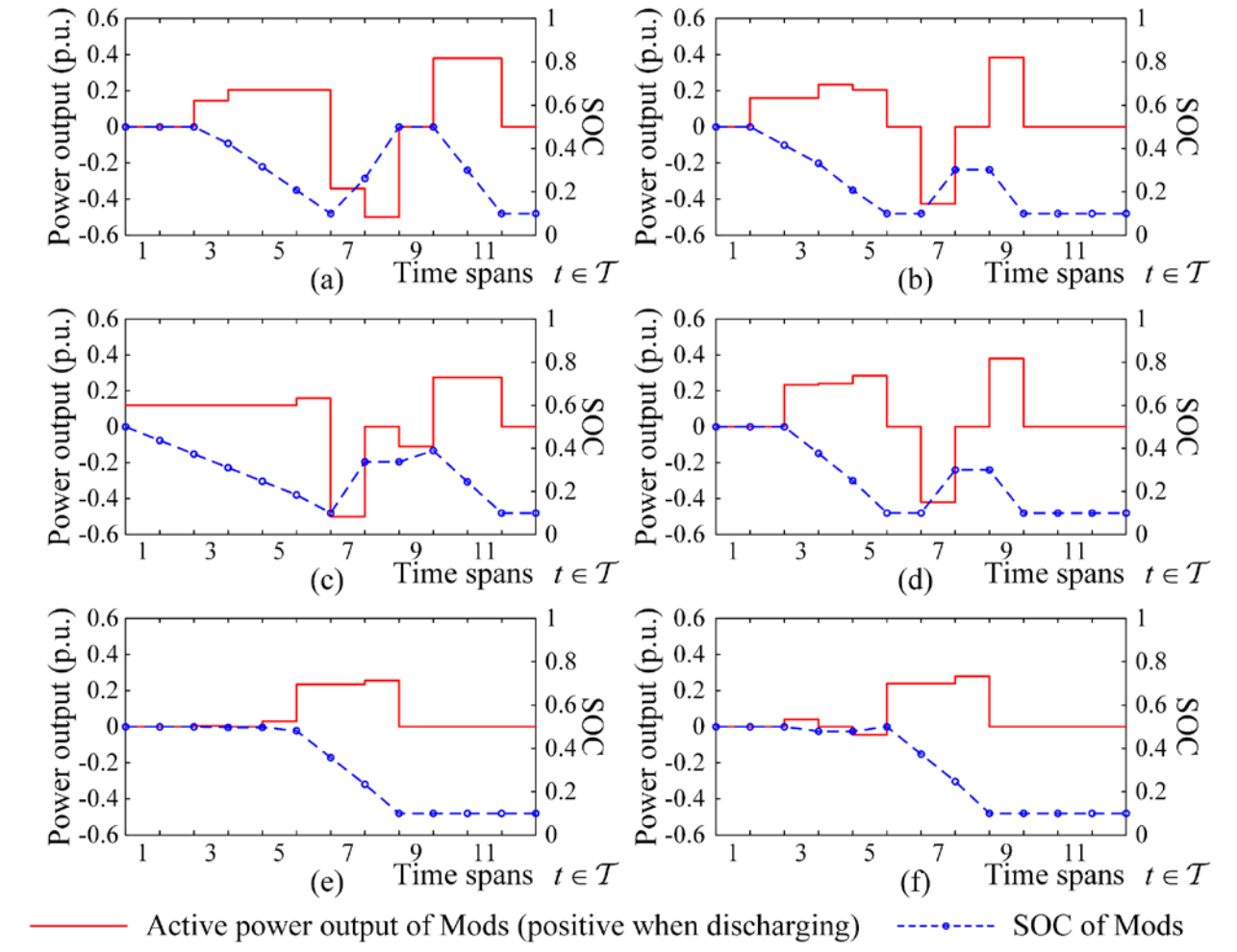

Fig. 14. Results of the active power output and SOC of (a) Mod 1, (b) Mod 2, (c) Mod 3, (d) Mod 4, (e) Mod 5, and (f) Mod 6.

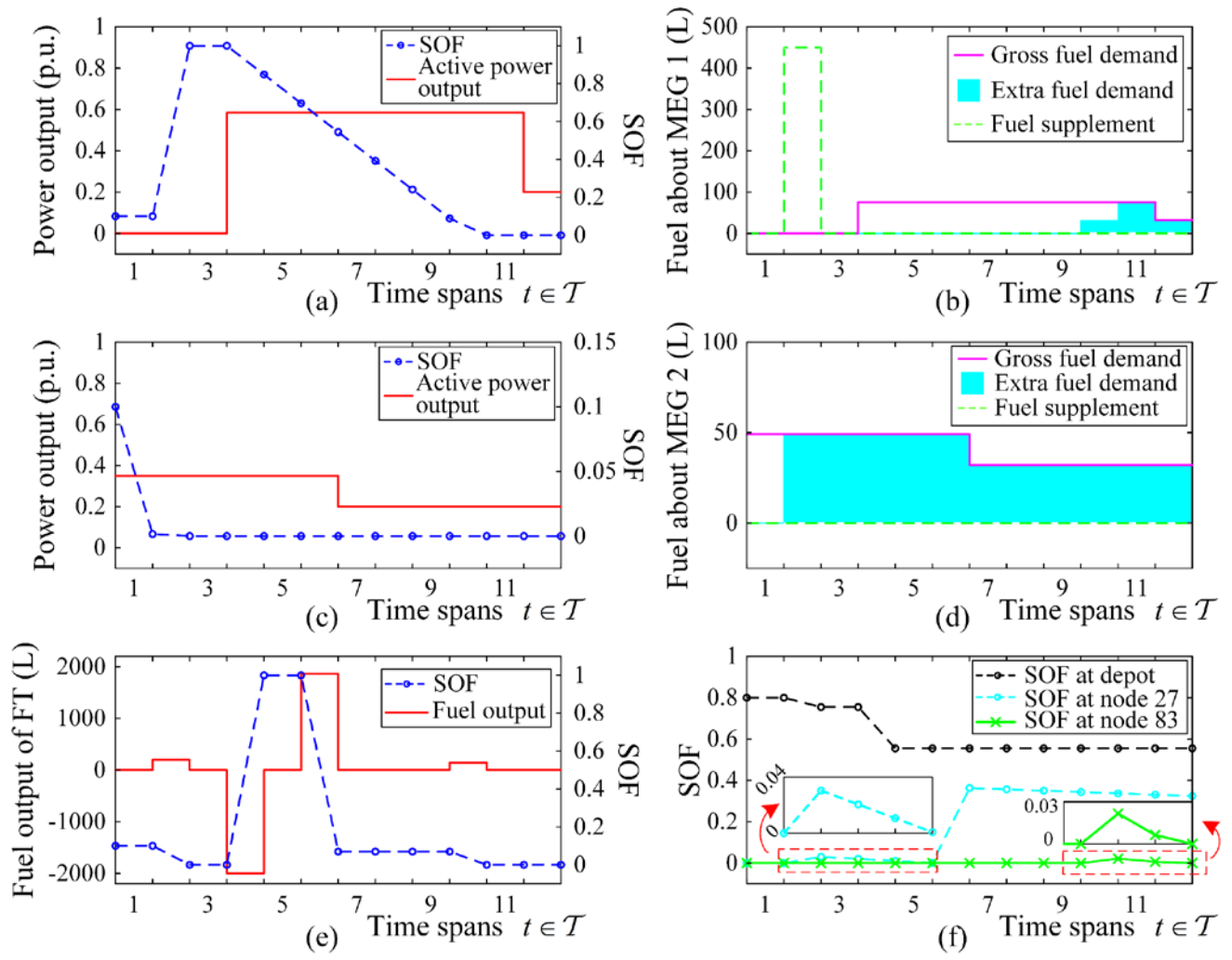

Fig. 15. Results of (a) active power output, SOF, (b) fuel demand/supplement of MEG 1; (c) active power output, SOF, (d) fuel demand/supplement of MEG 2; (e) fuel output and SOF of FT; and (f) SOF of nodes 27, 83 and the depot.

The main scheduling results are given in Fig. 12 to Fig. 15 and, from the perspectives of the several islands, they are described as follows.

For *Island 1*, the SMESSs are initially located there. When scheduling begins, all of them except Mod 3 are immediately carried away by the Carrs to other de-energized islands. Mod 3 stays and discharges continuously to supply the loads of *Island 1*, until *Island 1* regains the supply from the substation as the faulted branch (2, 8) is repaired after *t*=6. Then, Mod 3 charges intensively for one time span to restock itself and, in *t*=8, it is sent by Carr 1 to node 59 to supply the loads of *Island 5*.

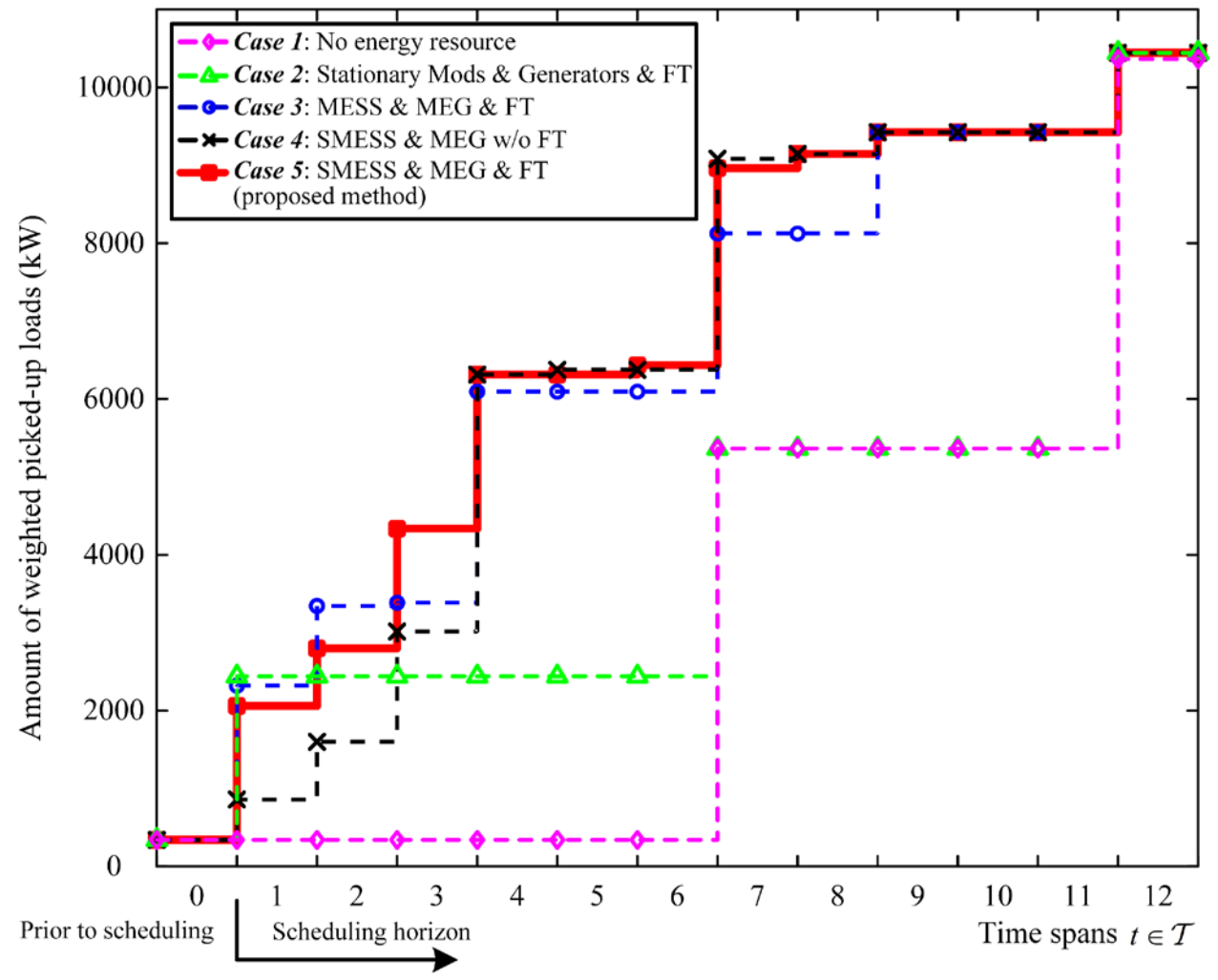

Fig. 16. Comparison among *Cases 1* to *5*.

For *Island 5*, it soon receives Mod 2 from *Island 1* in *t*=2, and Mod 2 supplies the loads of *Island 5* until its available stored energy is used up after *t*=5. Then, in *t*=6, Mod 2 is transported to *Island 1*, which will be reconnected to the substation after the arrival of Mod 2. After charging at *Island 1*, Mod 2, together with Mod 3, is carried back to *Island 5* in *t*=8. In addition, note that Mod 5 has already arrived at *Island 5* in *t*=5 and maintained the power supply to *Island 5* during the absence of Mod 2. Then, after *t*=8, *Island 5* is supplied by Mods 2, 3, and 5 until it is reconnected to the substation after *t*=11.

For *Island 3*, it obtains Mods 1 and 6 brought by Carr 2 in *t*=3. Then, Mod 1 works as the main power supply to *Island 3*, while Mod 6 is soon taken away by Carr 2 to node 104 of *Island 6*. After *t*=6, as some faulted branches are repaired, *Island 3* is reconnected to the substation and regains the normal supply. We see from Fig. 14 that Mod 1 charges intensively in *t*=7 and *t*=8. Then, Mod 1 is sent to *Island 6*, which is still isolated from the substation. In addition, Mod 4 is also carried to *Island 3* for energy supplement in *t*=7 and soon sent back to *Island 6*.

For *Island 6*, Mods 4 and 5 are brought by Carr 1 two time spans after scheduling begins, whereas Mod 5 is soon taken away by Carr 1 to *Island 5* after the arrival. Mod 4 provides the main supply to *Island 6* until its available stored energy is used up after *t*=5. Then, Mod 4 is sent to *Island 3* by Carr 2. After charging in *t*=7, Mod 4 is sent back to continue providing the power supply to *Island 6*. In addition, Mod 6, which arrives in *t*=5, discharges all available energy to maintain the supply to *Island 6* during the absence of Mod 4, *i.e.*, from *t*=6 to *t*=8. After *t*=8, *Island 6* is supplied alternately by Mods 4 and 1, the latter of which is brought by Carr 3 in *t*=9, until *Island 6* is reconnected to the substation.

For *Island 2*, MEG 2 stays at the initial location to supply *Island 2* during the whole blackout. Fuel for its generation is supplied twice by FT: When scheduling starts, FT immediately moves to *Island 2* even there is only a little fuel stored in it (an initial SOF of 0.1 is assumed for FT); then, FT returns to the depot to fully refuel itself in *t*=4 and, once again, it carries enough fuel to *Island 2* to maintain MEG 2's power output.

For *Island 7*, MEG 1 arrives and provides the supply after

$t$=3. Note that, as shown in Fig. 12 and Fig. 15, MEG 1 first moves to the depot to refuel itself after scheduling begins, and then MEG 1 moves to and supplies *Island 7*. From Fig. 15 (b), the fuel supplement at the depot is enough for a long operation of MEG 1. After $t$=9, extra fuel is required for MEG 1 and is supplied in time by FT.

In addition, note that a certain number of nodes in IEEE 123-node system have a zero load [34], including nodes 4, 66, etc., and this leads to useless results of whether they are picked up or not, such as the states of nodes 4 and 66 in Fig. 13. In other words, for node $i$ where the load is zero, the solution of $\delta_{i,t}$, though it is feasible for the model, truly makes no sense in practice and we can just ignore it.

*4) Case studies and Comparison*

Different cases are further studied for comparison as we did in Test A. *Cases 1* to *5* here are set in the same way as *Cases 1* to *5* in Test A, respectively, except some small details: In *Case 2*, we fix the six Mods and two MEGs at nodes 17 and 27, respectively; and in *Case 3*, three general MESSs, each with a 1 MW/2 MW·h capacity, are used instead of two in *Case 3* of Test A. The revisions to the model required to realize the above cases can be found in the appendix [33]. The results of comparison are given in Fig. 16, which verified the advantage of the proposed method over other measures in *Case 1* to *4*. Analysis about the comparison in detail can be similar as we described in Test A and, for space limit, is not described here.

## VI. Conclusion

In this paper, we propose the idea of SMESS to further evolve the scheduling of traditional MESS. In the concept of SMESS, a Carr is allowed to carry multiple Mods and the Carr and each of the Mods are scheduled as independent components, thus endowing the scheduling of SMESSs with more feasibility and flexibility than that of MESSs. The constraints for scheduling SMESSs are derived. The fuel delivery is modeled to schedule FTs to guarantee adequate fuel for MEGs operation. Then, aiming at the DS restoration issue, the joint scheduling of SMESSs, MEGs, FTs, and the DS reconfiguration is formulated as an MILP model. Numerical results demonstrate the effectiveness of the proposed method and its good performance in boosting DS resilience. The SMESS concept and the proposed joint scheduling model could result in a more effective and higher-level usage of mobile energy resources.